\documentclass[aps,prd,superscriptaddress,nofootinbib,11pt]{revtex4}
\usepackage[english]{babel}
\usepackage[utf8]{inputenc}
\usepackage{graphicx}   % need for figures
\usepackage{slashed}
\usepackage{epstopdf}
\usepackage{verbatim}   % useful for program listings
\usepackage{color}      % use if color is used in text
\usepackage{subfigure}  % use for side-by-side figures
\usepackage{multirow}
\usepackage{hyperref}   % use for hypertext links, including those to external documents and URLs
\usepackage{float}
\usepackage{epsfig,rotating}
\usepackage{amsmath,amssymb}
\usepackage{dsfont}
\restylefloat{table}
\numberwithin{equation}{section}

\newcommand{\vx}{\vec{x}}

\newcommand{\vk}{\vec{k}}

\newcommand{\be}{\begin{equation}}
\newcommand{\ee}{\end{equation}}
\newcommand{\bea}{\begin{eqnarray}}
\newcommand{\eea}{\end{eqnarray}}

\newcommand{\ta}{\widetilde{A}}
\newcommand{\tb}{\widetilde{B}}
\newcommand{\tf}{\widetilde{f}}
\newcommand{\ket}[1]{|#1\rangle}
\newcommand{\bra}[1]{\langle#1|}

\begin{document}
%\tableofcontents
\title{Non-adiabatic cosmological production of  ultra-light Dark Matter .}

\author{Nathan Herring}
\email{nmh48@pitt.edu} \affiliation{Department of Physics and
Astronomy, University of Pittsburgh, Pittsburgh, PA 15260}
\author{Daniel Boyanovsky}
\email{boyan@pitt.edu} \affiliation{Department of Physics and
Astronomy, University of Pittsburgh, Pittsburgh, PA 15260}
\author{Andrew R. Zentner}
\email{zentner@pitt.edu}
\affiliation{Department of Physics and
Astronomy, University of Pittsburgh, Pittsburgh, PA 15260}

 \date{\today}

\begin{abstract}
We study the non-adiabatic cosmological production of ultra light dark matter (ULDM) under a minimal set of assumptions: a free ultra light real scalar   as a  spectator field in its Bunch-Davies vacuum state during inflation and instantaneous reheating into a radiation dominated era. For (ULDM) fields minimally coupled to gravity, non-adiabatic particle production yields a \emph{distribution function} peaked at \emph{low} comoving momentum $\mathcal{N}_k \propto 1/k^3$. The infrared behavior is a remnant of the infrared enhancement of light minimally coupled fields during inflation.    We obtain the full energy momentum tensor, show explicity its  equivalence with the fluid-kinetic one in the adiabatic regime, and extract the abundance, equation of state and free streaming length (cutoff in the matter power spectrum). Taking the upper bound on the scale of inflation from Planck, the (UDLM) saturates the dark matter abundance for $m \simeq 1.5\,\times 10^{-5}\mathrm{eV}$ with an equation of state parameter $w \simeq 10^{-14}$ and a free streaming length $\lambda_{fs} \simeq 70\,\mathrm{pc}$. Thus this   cosmologically produced (ULDM) yields a \emph{cold} dark matter particle.  We argue that the abundance from non-adiabatic production yields a \emph{lower bound} on generic (ULDM) and axion-like particles that must be included in any assessment of (ULDM) as a dark matter candidate.

\end{abstract}


\maketitle

\section{Introduction}

Despite a large effort on the direct detection of weakly interacting massive particles in the mass range of few to $100\,\mathrm{GeV}$ with weak interactions cross sections, no particle beyond the Standard Model with these properties has been found\cite{bertone}-\cite{nowimp3}. This lack of evidence is  motivating the study of alternative \emph{light} or \emph{ultra-light} (DM) candidates, such as sterile neutrinos,  axions or axion-like particles, ``fuzzy'' dark matter (FDM), light dark scalars and dark vector bosons\cite{axionreviu}-\cite{dmpart}.  An (FDM) candidate with mass $m\simeq 10^{-22}\,\mathrm{eV}$, and  de-Broglie wavelength $\simeq \mathrm{kpc}$ could be a cold-dark matter (CDM) candidate with the potential for solving some small scale aspects of galaxy formation\cite{fuzzyDM}-\cite{jens}. All of these candidates are characterized by very small masses and couplings to Standard Model degrees of freedom. Lyman-$\alpha$\cite{Lyalfa,Lyalfaforest} and pulsar timing\cite{pulsarultralight} provide constraints on the mass range of (ultra) light dark matter (ULDM). Light dark matter (DM) candidates are not only probed by their gravitational properties\cite{DMgraviprobes} but there are various proposals for \emph{direct} detection,  from high energy colliders\cite{HECdarkfoton} to ``table-top'' experiments\cite{admx}-\cite{savas}. There are several proposed mechanisms of production of light or ultra-light dark matter\cite{axionreviu,axionsikivie,abbott,dine,hertzberg,dmreport,darksector,dmpart}.

Particle production in a dynamical cosmological background was studied in pioneering work in refs.\cite{parker,birrell,ford,fullbook,parkerbook,mukhabook}. Gravitational production of  (DM) candidates was studied for various candidates and within different settings: heavy (DM) particles\cite{heavydm1,heavydm2,heavydm3,kuzmin}, production from inflaton oscillations or oscillatory backgrounds \cite{vela,ema1,ema2}, for ``stiff'' equations of state in\cite{vilja2}, or during reheating\cite{hash,vilja1}. These previous studies considered heavy (DM) candidates and often invoked the adiabatic approximation valid for large masses and/or wavevectors.

In this article we study the gravitational production of ultra-light (DM) with important differences from  previous studies:

\textbf{i)} We study the \emph{non-adiabatic} gravitational production of\emph{ ultra-light dark matter} (ULDM) as a consequence of cosmological expansion during the inflationary and post-inflationary radiation dominated era until matter-radiation equality. We obtain the abundance, equation of state and free-streaming length (cutoff scale in the matter power spectrum) to assess whether this candidate describes cold, warm or hot (DM).

\textbf{ii)} We consider a real free scalar field describing the (ULDM) as a \emph{spectator field} during inflation, namely it does \emph{not} couple to the inflaton, it does \emph{not} acquire an expectation value, hence it does not contribute to \emph{linear} isocurvature perturbations that couple to long-wavelength metric perturbations\cite{gordon,byrnes,bartolo}. We discuss the issue of \emph{non-linear} entropy perturbations in section(\ref{sec:iso}). This scalar field is in its (Bunch-Davies) vacuum state during inflation. A vanishing expectation value of the field precludes a ``misalignement'' type production mechanism.

\textbf{iii)} This field does not feature self-interactions or interactions with any other field, it only interacts gravitationally.

 \textbf{iv)} We focus on scales that are well outside the horizon at the end of inflation, since these are the scales of cosmological relevance for structure formation, and  we  \emph{assume} a rapid transition from the inflationary stage to a radiation dominated (RD) era.

  \textbf{v)}  We obtain the full energy momentum tensor; its expectation value in the ``in'' Bunch-Davies vacuum yields the energy density and pressure. We show that in the asymptotic regime when the evolution becomes adiabatic, the zeroth-order adiabatic energy momentum tensor coincides with the usual fluid-kinetic one. We obtain the abundance, equation of state and free-streaming length near matter-radiation equality to assess whether this candidate describes cold, warm or hot (DM). Imposing the observed (DM)  abundance yields a bound on the mass of the (ULDM) particle which \emph{only} depends on cosmological parameters.

 We discuss (ULDM) minimally and conformally coupled to gravity. Although we expect negligible production of an (ULDM) particle conformally coupled to gravity, its detailed study provides an explicit quantitative confirmation of this expectation and highlights the main differences with the case of minimal coupling. The comparison between the minimally and conformally coupled cases allow us to conclude that in the minimal coupling scenario, substantial particle production occurs during inflation after the corresponding wavelengths become super-horizon.

 \textbf{Summary of main results:} For a minimally coupled light scalar field taken as \emph{spectator} in its Bunch-Davies vacuum state during inflation, non-adiabatic particle production yields a \emph{distribution function} peaked at small comoving momentum $\mathcal{N}_k \propto 1/k^3$. The low momentum enhancement is a distinct remnant of the infrared enhancement of light minimally coupled fields during inflation. Assuming the upper bound on the scale of inflation established by Planck\cite{planck2018}, we find that a mass $\simeq 10^{-5}\,\mathrm{eV}$ yields the correct dark matter abundance. Furthermore, we find that this (DM) candidate, despite being very light is extremely cold; its equation of state parameter at matter-radiation equality is $w\simeq 10^{-14}$ and features a free streaming length (cutoff scale in the matter power spectrum) $\lambda_{fs} \simeq 70\,\mathrm{pc}$. Conformally coupled (ULDM) features a negligible abundance.

 The results of this study apply also to axion-like particles, albeit with no other interactions but gravitational.    The abundance, equation of state, and clustering  properties only depend on cosmological parameters and the mass, therefore this study provides the simplest scenario for particle production of (ULDM), and for a long-lived (DM) candidate a \emph{lower bound} on the abundance. This lower bound on the abundance from non-adiabatic cosmological production should enter in \emph{any} assessment of (ULDM) candidates, even those with interactions.

 \vspace{1mm}

 The model of (ULDM) is introduced in section (\ref{sec:model}). Section (\ref{sec:asy}) discusses the   ``in'' states and define the  ``out'' particle states, obtaining the number of asymptotic ``out'' particles produced non-adiabatically for minimal and conformal coupling to gravity. In section (\ref{sec:nonad}) we discuss the non-adiabatic nature of particle production.  Section (\ref{sec:tmunu}) analyzes the energy momentum tensor, discusses renormalization aspects, establishes the relation with the fluid-kinetic energy momentum tensor in the adiabatic regime and defines the energy density and pressure of the asymptotic particle states. In this section we obtain the relation between the dark matter abundance, the particle's mass and cosmological parameters. We also obtain the equation of state and free-streaming length and establish that non-adiabatic production of (ULDM) yields a \emph{cold dark matter } candidate. In section (\ref{sec:iso}) we discuss linear and non-linear entropy perturbations. Section (\ref{sec:discussion}) discusses various aspects and caveats suggesting further questions and avenues of study, and section (\ref{sec:conclusions}) summarizes our conclusions. Two appendices provide technical details.

 \vspace{1mm}

\section{The model for the (ULDM) scalar. }\label{sec:model}

We consider a free real ultra-light scalar degree of freedom as a dark matter candidate (ULDM) and invoke the following main assumptions:

\textbf{i:)} It is  a  \emph{spectator} field during inflation. Namely, it  does not interact with any other field, including the inflaton, and it does not acquire a vacuum expectation value, therefore it does not drive inflation. Because it does not acquire an expectation value  it does not contribute to \emph{linear} isocurvature perturbations that source long-wavelength metric perturbations\cite{gordon,byrnes,bartolo}. See section (\ref{sec:iso}) for a discussion on \emph{non-linear} entropy perturbations.

\textbf{ii:)} The inflationary stage is described by an exact de Sitter space-time,   the ultralight field is in the Bunch-Davies vacuum state and we consider field fluctuations with superhorizon wavelengths at the end of inflation, since these are the wavelengths of cosmological relevance for structure formation.

\textbf{iii:)} We assume instantaneous reheating: namely we consider an instantaneous transition from the inflationary to a radiation dominated stage post-inflation. There is as yet an incomplete understanding of the non-equilibrium dynamics of reheating. Reheating dynamics  depend  crucially on various assumptions on couplings with the inflaton and/or other fields, and thermalization processes\cite{reheat} in an expanding cosmology. The question of how the nearly $\simeq 100$ degrees of freedom of the Standard Model attain a state of local thermodynamic equilibrium after inflation and on what time scales is still unanswered.   Most studies \emph{model} the couplings and dynamics; therefore  any model of reheating is at best tentative and very approximate.  We bypass the inherent ambiguities and model dependence of the reheating dynamics, and  assume \emph{instantaneous} reheating after inflation  to a radiation dominated (RD) era.  The physical reason behind this assumption is that we are primarily concerned with wavevectors that have crossed the Hubble radius during inflation well before the transition to (RD) and are well outside the horizon during this transition, hence causally decoupled from microphysics. These modes feature very slow dynamics at the end of inflation, and the assumption that they are frozen during the reheating time interval seems physically warranted (see further discussion in section (\ref{sec:discussion})).  We assume that both the scale factor and the Hubble rate are \emph{continuous} across the transition. Along with the continuity of the mode functions and their time derivative across the transition (see below), this, in fact, entails the continuity of the energy density obtained from the energy momentum tensor (see below).

 \textbf{iv:)} Unlike previous studies that invoked the adiabatic approximation, we study \emph{non-adiabatic} cosmological production of (ULDM). This is a direct consequence of a very small mass and  field fluctuations with superhorizon wavelengths after inflation.

 \textbf{v:)} The (RD) era is dominated by a large number $\simeq 100$ of ultrarelativistic degrees of freedom justifying  taking the space time metric  during this era as a \emph{background} and neglecting the contribution from the single scalar degree of freedom.

In comoving
coordinates, the action for the real (ULDM) scalar field is given by
\be
S   =  \int d^3x \; dt \;  \sqrt{-g} \,\Bigg\{
\frac{1}{2}{\dot{\phi}^2}-\frac{(\nabla
\phi)^2}{2a^2}-\frac{1}{2} \, \big[m^2 +\xi \,R\big]  \phi^2 \Bigg\} \label{lagrads}
\ee where
\be R= 6\Big[\frac{\ddot{a}}{a}+\Big(\frac{\dot{a}}{a} \Big)^2 \Big] \,,\label{ricci}\ee is the Ricci scalar, (here the dot stands for derivatives with respect to comoving time $t$)  and $\xi$ is the coupling  to gravity, with $\xi=0, 1/6$ corresponding to minimal or conformal coupling, respectively, we will study both cases separately. We consider a spatially flat Friedmann-Robertson-Walker (FRW) cosmology in conformal time coordinate, with the metric given by

\be
g_{\mu\nu}= a^2(\eta) \;  \eta_{\mu\nu}
 \,,\label{gmunu}
\ee

\noindent where  $\eta_{\mu\nu}=\textrm{diag}(1,-1,-1,-1)$ is the flat
Minkowski space-time metric.

Introducing the conformally rescaled fields
\be   \phi(\vx,t) = \frac{\chi(\vx,\eta)}{a(\eta)} \,, \label{rescaledfields}\ee  with
\be R= 6 \, \frac{a''(\eta)}{a^3(\eta)}\,, \label{Ricciconformal} \ee the primes now refer to derivatives with respect to conformal time. The action becomes (neglecting an irrelevant surface term that does not affect the equations of motion or energy momentum tensor),

 \be  S    =
  \int d^3x \; d\eta \, \frac12\left[
{\chi'}^2-(\nabla \chi)^2- \mathcal{M}^2 (\eta) \; \chi^2 \right] \,, \label{lagchi}\ee where

 \be
  \mathcal{M}^2(\eta) =m^2 \,a^2(\eta)- \frac{a''(\eta)}{a(\eta)}\,(1-6\xi)  \,.  \label{massconformal} \ee

The inflationary stage is  described by a   spatially flat de Sitter space time (thereby neglecting slow roll corrections)  with a scale factor
\be a(\eta) = -\frac{1}{H_{dS}(\eta-2\eta_R)} \,,\label{adS} \ee where $H_{dS}$ is the Hubble constant during de Sitter and $\eta_R$ is the (conformal) time at which the de Sitter stage transitions to the (RD) stage.

During the   radiation dominated (RD)  stage  the scale factor is given by
\be a(\eta)=  H_R\,\eta   \label{Crdmd}\ee with
\be H_R= H_0\,\sqrt{\Omega_R}\simeq 10^{-35}\,\mathrm{eV}   \,,  \label{Hs}\ee and matter radiation equality occurs at
\be a_{eq}= \frac{\Omega_R}{\Omega_M} \simeq 1.66\,\times 10^{-4}  \,.\label{rands}\ee

We model the transition from de Sitter to (RD) at a (conformal) time $\eta_R$  by requiring that the scale factor and the Hubble rate be continuous across the transition at   $\eta_R$,   assuming self-consistently  that the transition occurs deep in the (RD) era so that $a(\eta_R) = H_R\,\eta_R \ll a_{eq}$. Continuity of the scale factor and Hubble rate at the instantaneous reheating time results in that the energy density, namely the expectation value of  $T^{0}_0$ is \emph{continuous at the transition}. This important aspect is discussed further in section (\ref{sec:tmunu}).

 Using   $H(\eta) = a'(\eta)/a^2(\eta)$,  continuity of the scale factor and Hubble rate at $\eta_R$  imply that
\be a_{dS}(\eta_R) = \frac{1}{H_{dS}\,\eta_R}= H_R\,\eta_R ~~;~~ H_{dS} = \frac{1}{H_R\,\eta^2_R} \,, \label{transition} \ee yielding
\be \eta_R = \frac{1}{\sqrt{H_{dS}\,H_R}}\,.  \label{etaR}\ee

The most recent constraints from Planck on the tensor-to-scalar ratio is\cite{planck2018}

\be H_{dS}/M_{Pl} < 2.5\times 10^{-5} ~~(95\%)\,\mathrm{CL} \,. \label{planckcons}\ee We take as a representative value $H_{dS} = 10^{13}\,\mathrm{GeV}$, from which it follows that
\be a_{dS}(\eta_R) =H_R\,\eta_R = \sqrt{\frac{H_R}{H_{dS}}} \simeq 10^{-28}\ll a_{eq} \,. \label{scalefac}\ee This scale corresponds to an approximate ambient radiation temperature after the transition from de Sitter to (RD)
\be T(\eta_R) \simeq \frac{T_0}{a_{RD}(\eta_R)} \simeq 10^{15}\,\mathrm{GeV} \label{Tgut}\ee where $T_0\propto 10^{-4}\,\mathrm{eV}$ is the CMB temperature today.

 We also define the mass of the (DM) particle in units of $\mathrm{eV}$ as
\be  m_{ev}\equiv \frac{m}{(\mathrm{eV})} \,,  \label{mev}\ee  which for ultra-light (DM) particles  we define as  $m_{ev} \ll 1$.

\section{Asymptotics: ``in-out'' states,  adiabatic mode functions  and particle states. }\label{sec:asy}

\subsection{Asymptotic ``in-out'' states.}

The quantization of the real (ULDM)  scalar field in a finite comoving volume $V$ proceeds by writing
\be \chi(\vx,\eta) = \frac{1}{\sqrt{V}}\,\sum_{\vk} \Big[ a_{\vk}\,g_k(\eta)\,e^{-i\vk\cdot\vx} + a^\dagger_{\vk}\,g^*_k(\eta)\,e^{i\vk\cdot\vx}\Big]\,, \label{quantchi}\ee where $\vk$ are comoving wave vectors. The mode functions $g_k(\eta)$  are solutions of  the equations of motion

\be
  g''_{k}(\eta)+
\Big[k^2+m^2 \,a^2(\eta)- \frac{a''(\eta)}{a(\eta)}\,(1-6\xi)
\Big] g_{k}(\eta)  =   0\,,    \label{chimodes}\ee   and  are normalized to obey the Wronskian condition
 \be g'_k(\eta)\,g^*_{k}(\eta) - g_k(\eta)\,{g'}^{*}_k(\eta) = -i \label{wronskian}\ee so that
 $a_{\vk},a^\dagger_{\vk}$ obey canonical commutation relations.

 A familiar interpretation of the mode equation follows by writing (\ref{chimodes}) as
 \be -\frac{d^2}{d\eta^2}\, g_k(\eta) + V(\eta)g_k(\eta) = k^2 g_k(\eta) ~~;~~ V(\eta) = -m^2 a^2(\eta) + (1-6\xi)\,\frac{a''(\eta)}{a(\eta)} \,,\label{scheqn}\ee namely a Schroedinger   equation for a wave function $g_k$ with a potential $V(\eta)$ and ``energy'' $k^2$. The potential $V(\eta)$ and/or its derivative are discontinuous at the transition $\eta_R$;  however  $g_k(\eta)$  and $g'_k(\eta)$ are continuous at $\eta_R$. Defining
 \be g_k(\eta) = \Bigg\{ \begin{array}{c}
                   g^<_k (\eta) ~~; ~~ \mathrm{for} ~~;~~ \eta < \eta_R\\
                   g^>_k (\eta) ~~; ~~ \mathrm{for} ~~;~~ \eta > \eta_R\\
                 \end{array} \,,\label{geta} \ee  the matching conditions are
\bea g^<_k(\eta_R) & = & g^>_k(\eta_R) \nonumber \\
\frac{d}{d\eta} g^<_k(\eta)\Big|_{\eta_R} & = &     \frac{d}{d\eta} g^>_k(\eta)\Big|_{\eta_R} \,.    \label{matchcond} \eea

As is discussed below  (see section (\ref{sec:tmunu})), these continuity conditions on the mode functions, along with the continuity of the scale factor and Hubble rate at the transition   ensures that the energy density is \emph{continuous} at the transition from inflation to (RD).

 \subsubsection{Inflationary stage:}

 \vspace{1mm}

We consider that the (ULDM) scalar is in the Bunch-Davies vacuum state during the inflationary stage, which corresponds to the mode functions $g_k(\eta)$ fulfilling the boundary condition
\be g_k(\eta) ~~~ _{ \overrightarrow{\eta \rightarrow -\infty}} ~~~ \frac{e^{-ik\eta}}{\sqrt{2k}} \,, \label{BD} \ee and the Bunch-Davies vacuum state $\ket{0}$ is such that
\be a_{\vk}\ket{0}=0  \, \, \forall  \vk \,.\label{BDvac} \ee  We refer to this vacuum state as the \emph{in} vacuum.

We will consider both cases: conformal coupling (CC) $\xi =1/6$ and minimal coupling (MC) $\xi =0$.

During the de Sitter stage ($\eta < \eta_R$), with the scale factor given by eqn. (\ref{adS}), the mode equation becomes
\be \frac{d^2}{d\tau^2}\,g^{<}_k(\tau)+\Big[ k^2 - \frac{\nu^2-1/4}{\tau^2}\Big]\,g^{<}_k(\tau) =0 \,, \label{modeqndS}\ee where
\be \tau = \eta-2\eta_R~~;~~ \nu^2 = \Bigg\{  \begin{array}{c}
                                       \frac{9}{4}-\frac{m^2}{H^2_{dS}}~~\mathrm{for}~~\xi =0 ~~ (MC) \\
                                       \frac{1}{4}-\frac{m^2}{H^2_{dS}} ~~\mathrm{for}~~\xi =1/6  ~~ (CC)
                                     \end{array} \Bigg\}
                                     \,. \label{dSparas}\ee The solution with the boundary
condition (\ref{BD}) is given by

\be g^{<}_k(\tau) = \frac{1}{2}\,\sqrt{-\pi \tau}\,\, e^{i\frac{\pi}{2}(\nu+1/2)}\,H^{(1)}_\nu(-k\tau) \label{BDsolution}\ee where $H^{(1)}_\nu$ is a Bessel function. We note that with   $H_{dS} \simeq 10^{13}\,\mathrm{GeV}$ it follows that $m/H_{dS} \simeq m_{ev}\,10^{-22} \ll 10^{-22}$ and can be safely ignored in the expression for $\nu$. Therefore, neglecting the mass of the (ULDM) scalar, we find
\be g^{<}_k(\tau) = \Bigg\{ \begin{array}{c}
 \frac{e^{-ik\tau}}{\sqrt{2k}}\,\Big[ 1-\frac{i}{k\tau}\Big]~~\mathrm{for}~~\xi=0 ~~(MC)\\

                          \frac{e^{-ik\tau}}{\sqrt{2k}}~~\mathrm{for}~~\xi=1/6 ~~ (CC)

                        \end{array}
\Bigg\}\,.  \label{BDsols2}\ee

With $H_{dS} \simeq 10^{13}\,\mathrm{GeV}$ we find that $\eta_R \simeq   10^{6}\,\mathrm{eV}^{-1} \simeq 0.2\,\mathrm{meters}$. In what follows we will consider that all the modes of cosmological interest are \emph{well outside} the Hubble radius at the end of inflation, namely
\be k\,\eta_R \ll 1\,,  \label{superhori}\ee  for the value of $H_{dS}$ assumed above, with $\eta_R \simeq 10^{6}\,(\mathrm{eV})^{-1}$ the superhorizon condition (\ref{superhori}) corresponds to comoving wavevectors $k \ll \mu\mathrm{eV}$ or comoving wavelengths $\gg 1 ~\mathrm{meters}$, obviously including \emph{all} astrophysically relevant scales.

  The ``\emph{in}''  state is the Bunch-Davies vacuum  defined by equation (\ref{BDvac}) and the mode functions (\ref{BDsols2}) during the   inflation stage, taken to be de Sitter space-time, thereby neglecting small slow-roll corrections.

\subsubsection{Radiation   dominated era:}

During the radiation era  for $\eta > \eta_R$, with $a(\eta) = H_R\eta$    we set $a''=0$, and the mode equation (\ref{chimodes})  becomes
\be  \frac{d^2}{d\eta^2}g^{>}_k(\eta)+\Big[k^2+m^2 \,H^2_R \,\eta^2\Big]g^{>}_k(\eta) =0 \,, \label{paracyl}\ee the general solutions of which are linear combinations of  parabolic cylinder functions\cite{as,nist,bateman,magnus}. As ``out'' boundary conditions, we impose that such a combination should describe asymptotically positive frequency ``particle'' states and their hermitian conjugate. This identification relies on a WKB form of the asymptotic mode functions.

 Let us  consider a particular solution of (\ref{paracyl}) of the WKB form
\be f_k(\eta) = \frac{e^{-i\,\int^{\eta}_{\eta_R}\,W_k(\eta')\,d\eta'}}{\sqrt{2\,W_k(\eta)}} \,. \label{WKB}\ee Upon inserting this ansatze in the mode equation (\ref{paracyl}) one finds that $W_k(\eta)$ obeys
\be W^2_k(\eta)= \omega^2_k(\eta)- \frac{1}{2}\bigg[\frac{W^{''}_k(\eta)}{W_k(\eta)} - \frac{3}{2}\,\bigg(\frac{W^{'}_k(\eta)}{W_k(\eta)}\bigg)^2 \bigg]\,,  \label{WKBsol} \ee where
\be \omega^2_k(\eta) = k^2+m^2 \,H^2_R \,\eta^2\,. \label{omegak}\ee

 When $\omega_k(\eta)$ is a slowly-varying function of time the WKB eqn. (\ref{WKBsol}) may be solved in a consistent \emph{adiabatic expansion} in terms of derivatives of $\omega_k(\eta)$ with respect to $\eta$ divided by appropriate powers of the frequency,  namely
\be W^2_k(\eta)= \omega^2_k(\eta) \,\bigg[1 - \frac{1}{2}\,\frac{\omega^{''}_k(\eta)}{\omega^3_k(\eta)}+
\frac{3}{4}\,\bigg( \frac{\omega^{'}_k(\eta)}{\omega^2_k(\eta)}\bigg)^2 +\cdots  \bigg] \,.\label{adexp}\ee  We refer to terms that feature $n$-derivatives of $\omega_k(\eta)$ as of n-th adiabatic order.  During the time interval of rapid variations of the frequencies the concept of particle is ambiguous, but at long time the frequencies evolve slowly   and the concept of particle becomes clear.

 We want to identify ``particles'' (dark matter ``particles'') near the time of matter radiation equality, so that entering in the matter dominated era we can extract the energy density and pressure (energy momentum tensor) associated with dark matter\emph{ particles}. Therefore, we seek to clearly define the concept of particles near matter-radiation equality namely $a(\eta) \simeq a_{eq} \simeq 10^{-4}$.

  The condition of adiabatic expansion  relies on the ratio
\be \frac{\omega^{'}_k(\eta)}{\omega^{2}_k(\eta)} \ll 1 \,. \label{adiacondi}\ee  An upper bound on this ratio  is obtained in the very long wavelength (superhorizon) limit, taking $\omega_k(\eta) = m \,a(\eta)$,  in an (RD) cosmology  leads to the condition
\be \frac{a'(\eta)}{m \, a^2(\eta)} = \frac{H_R}{m\,a^2(\eta)}  \ll 1  \Longrightarrow a(\eta) \gg \frac{10^{-17}}{\sqrt{m_{ev}}} \,.  \label{alaful}\ee  Therefore, even for  $m_{ev}\simeq 1$ corresponding to to $a(\eta) \simeq 10^{-17}$ there is a long period of \emph{non-adiabatic} evolution  since the end of inflation $a(\eta_R) \simeq 10^{-29}  \ll 10^{-17}/\sqrt{m_{ev}}$, during which the $\omega_k(\eta)$ varies \emph{rapidly}. However, even for an  ultra-light particle with $m_{ev} \simeq 10^{-22}$  yielding  a much longer period of non-adiabatic evolution,  the adiabatic condition is  fulfilled well before matter-radiation equality. The adiabaticity condition becomes less stringent for non-vanishing wavevectors with $k \gg m\,a(\eta)$.

In conclusion,  the evolution of the mode functions becomes adiabatic  well before matter radiation equality. During  the adiabatic regime the WKB mode function (\ref{WKB}) asymptotically becomes
\be f_k(\eta) \rightarrow  \frac{e^{-i\,\int^{\eta}\,\omega_k(\eta')\,d\eta'}}{\sqrt{2\,\omega_k(\eta)}}\,, \label{outstates}\ee we refer to the mode functions with this asymptotic boundary condition as ``out''  particle states which obey the Wronskian condition
\be  f^{'}_k(\eta)\, f^{*}_k(\eta) - f_k(\eta)\,f^{' *}_k(\eta) = -i \,.\label{wronskf}\ee

The definition of these mode functions as describing particle states merits discussion. Our space time is \emph{not} Minkowski space time; dark energy entails that the cosmology describing our space time is nearly de Sitter (if dark energy is in the form of a cosmological constant), and Minkowski space time is a local approximation valid on scales much smaller than the Hubble scale.  The conformal and (local) comoving energy are related by
\be \omega_k(\eta) = \sqrt{k^2+m^2 a^2(\eta)} = a(\eta) \, E_k(\eta)\,, \label{omega} \ee with
\be E_k(\eta) = \sqrt{k^2_{ph}(\eta)+m^2} ~~;~~ k_{ph}(\eta) \equiv \frac{k^2}{a^2(\eta)} \,, \label{energy} \ee where $k_{ph}(\eta)$ is the physical momentum.

Consider the asymptotic phase of the mode function $f_k(\eta)$ given by eqn. (\ref{outstates}), using the relations (\ref{omega}, \ref{energy}) and $a(\eta) \,d\eta = dt$ with $t$ being cosmic time, it follows that
\be \int^{\eta}_{\eta_0} \omega_k(\eta')\,d\eta' = \int^{t}_{t_0} E_k(t')\, dt' \,. \label{expof}\ee Expanding around the lower limit and integrating we find
\be \int^{t}_{t_0} E_k(t')\, dt' = E_k(t_0)\,(t-t_0)\,\Bigg[1 - \frac{ 1 }{2}\,\beta^2_k(t_0) \, H(t_0)\,(t-t_0) + \cdots \Bigg] \,,\label{Eexpa} \ee where
\be \beta_k(t_0) = \frac{k_{ph}(t_0)}{E_k(t_0)} ~~;~~ H(t_0) = \frac{\dot{a}(t_0)}{a(t_0)}\,, \label{Expa2}\ee with $H(t_0)$ the Hubble expansion rate at $t_0$. Therefore it is clear that the phase is associated with particle states over a time scale $t-t_0 \ll 1/H(t_0) \simeq 13 \,\mathrm{Gyr}$. Thus on these time scales Minkowski space-time particle states are a valid description. This, of course is just a consequence of the equivalence principle.

The general solution of equation (\ref{paracyl}) is
\be g^>_k(\eta) = A_k\,f_k(\eta) + B_k \,f^*_k(\eta)\,,  \label{generalsol} \ee where $f_k(\eta)$ are the solutions of the mode equation (\ref{paracyl})   with asymptotic boundary conditions (\ref{outstates}) and $A_k$ and $B_k$ are Bogoliubov coefficients.  Since $g^>_k(\eta)$ obeys the Wronskian condition (\ref{wronskian}) and so does $f_k(\eta)$, it follows that the Bogoliubov coefficients obey
\be |A_k|^2 - |B_k|^2 =1 \,. \label{condiAB}\ee

Using the Wronskian condition (\ref{wronskf}) and the matching condition (\ref{matchcond}), we find that the Bogoliubov coefficients are determined from the following relations,

\bea A_k & = & i \Big[g^{'\,<}_k(\eta_R)\,f^*_k(\eta_R) - g^{<}_k(\eta_R)\,f^{'\,*}_k(\eta_R)\Big]  \nonumber \\
B_k & = & -i \Big[g^{'\,<}_k(\eta_R)\,f_k(\eta_R) - g^{<}_k(\eta_R)\,f^{'}_k(\eta_R)\Big] \,. \label{ABcoefs}\eea Since the mode functions $g^<_k(\eta)$  also fulfill the Wronskian condition (\ref{wronskian}), it is straightforward to confirm the identity (\ref{condiAB}).

For $\eta > \eta_R$ the field expansion (\ref{quantchi}) yields
\be \chi(\vx,\eta) = \frac{1}{\sqrt{V}}\,\sum_{\vk} \Big[ a_{\vk}\,g^>_k(\eta)\,e^{-i\vk\cdot\vx} + a^\dagger_{\vk}\,g^{*\,>}_k(\eta)\,e^{i\vk\cdot\vx}\Big] = \frac{1}{\sqrt{V}}\,\sum_{\vk} \Big[ b_{\vk}\,f_k(\eta)\,e^{-i\vk\cdot\vx} + b^\dagger_{\vk}\,f^{*}_k(\eta)\,e^{i\vk\cdot\vx}\Big] \,, \label{quantchig}\ee where
\be b_{\vk} = a_k \,A_k + a^{\dagger}_{-\vk}\,B^*_k ~~;~~ b^{\dagger}_{\vk} = a^{\dagger}_{\vk}\,A^*_k + a_{-\vk}\,B_k \,. \label{bops} \ee We refer to $b_{\vk},b^\dagger_{\vk}$ as the annihilation and creation operators of \emph{out particle} states respectively. They obey canonical quantization conditions as a consequence of the relation (\ref{condiAB}). In the Heisenberg picture the field operators evolve in time but the states do not. The vacuum state $|0\rangle$ is the Bunch-Davies vacuum state (\ref{BDvac}) in which the number of \emph{out}-particles is given by
\be \mathcal{N}_k = \bra{0}b^\dagger_{\vk} b_{\vk}\ket{0} = |B_k|^2\,.  \label{number}\ee We identify $\mathcal{N}_k$ with the number of dark matter particles produced \emph{asymptotically} from cosmic expansion. Only in the asymptotic adiabatic regime   can $\mathcal{N}_k$ be associated with the number of \emph{particles}. This point will be discussed further in section (\ref{sec:discussion}).

It remains to obtain the solutions $f_k(\eta)$ of the mode equations (\ref{paracyl}) with asymptotic ``out'' boundary condition (\ref{outstates}) describing asymptotic particle states.

It is convenient to introduce the dimensionless variables
\be x = \sqrt{2mH_R}\, \eta ~~;~~ \alpha = -\frac{k^2}{2mH_R}\,, \label{weberparas} \ee in terms of which the equation (\ref{paracyl}) is identified with Weber's equation\cite{as,nist,bateman,magnus}
\be \frac{d^2}{dx^2}\,g(x) +\Big[\frac{x^2}{4}-\alpha \Big]g(x) =0 \label{webereq}\ee
whose real solutions are Weber's parabolic cylinder functions\cite{nist,as,bateman,magnus}:
\be W[\alpha; \pm x] = \frac{1}{2^{3/4}}\,\Bigg[\sqrt{\frac{G_1}{G_3}}\,\,Y_1(\alpha;x)\mp \sqrt{\frac{2\,G_3}{G_1}}\,\,Y_2(\alpha;x) \Bigg]\,,\label{Wfunctions}\ee where
\be G_1 = \Bigg|\Gamma\Big(\frac{1}{4}+i\frac{\alpha}{2}\Big)\Bigg| ~~;~~ G_3 = \Bigg|\Gamma\Big(\frac{3}{4}+i\frac{\alpha}{2}\Big)\Bigg| \label{Gs}\ee and\cite{as,nist}
\bea Y_1(\alpha;x) & = & 1+ \alpha\,\frac{x^2}{2 !} + \big(\alpha^2 -\frac{1}{2}\big)\,\frac{x^4}{4!}+\cdots \label{Y1}\\
Y_2(\alpha;x) & = & x\Big[ 1+ \alpha\,\frac{x^2}{3 !} + \big(\alpha^2 -\frac{3}{2}\big)\,\frac{x^4}{5!}+\cdots \Big]\,.\label{Y2}\eea With these real solutions we construct the complex solution that satisfies the Wronskian condition (\ref{wronskf}) and features the asymptotic ``out-state'' behavior (\ref{outstates}) with $\omega^2_k(\eta) = \frac{x^2}{4}-\alpha$. It is straightforward to confirm that such a solution is given by (see appendix (\ref{app:weber}))

\be f_k(\eta) = \frac{1}{(8mH_R)^{1/4}}\,\Big[\frac{1}{\sqrt{\kappa}}\,W[\alpha;x] -i\sqrt{\kappa}\,W[\alpha;-x]  \Big]~~;~~ \kappa = \sqrt{1+e^{-2\pi|\alpha|}}-e^{-\pi|\alpha|}\,.\label{fconf}\ee It is shown in appendix (\ref{app:weber}) that these solutions do indeed satisfy the asymptotic ``out'' boundary condition (\ref{outstates}) and fulfill the Wronskian condition (\ref{wronskf}).

 The Bogoliubov coefficients are obtained from eqns. (\ref{ABcoefs}), where   the mode functions during the de Sitter era $g^<_k(\eta)$ are given by eqn. (\ref{BDsols2}) (with $\tau = \eta-2\eta_R$).

For $\eta_R= 1/\sqrt{H_{dS}\,H_R}$ (see eqn. (\ref{etaR})) it follows that
\be x(\eta_R) = \sqrt{\frac{2m}{H_{dS}}} \simeq \sqrt{2\,m_{ev}}\,\,10^{-11}~~;~~ -\alpha\,x^2(\eta_R) = (k\,\eta_R)^2 \ll 1 \,, \label{atetaR}\ee therefore for $\eta \simeq \eta_R$ we can set $Y_1(x) \simeq 1;Y_2(x) \simeq x$ in order to obtain the Bogoliubov coefficients from equation (\ref{ABcoefs}).

We note that the condition $x(\eta) \ll 1$ implies that
\be \frac{1}{m\,H_R\,\eta^2} = \frac{a'(\eta)}{m\,a^2(\eta)} \gg 1\,. \label{nonadiab}\ee Therefore, comparing with the   condition for adiabaticity (\ref{alaful}) we see that the mode functions after the transition are strongly non-adiabatic.

The regime of \emph{non-adiabatic} evolution is where particle production is most effective (see discussion in section (\ref{sec:discussion})). Furthermore, particle production is enhanced at \emph{longer wavelengths} because these modes feature the strongest departure from adiabaticity.

We emphasize that while we assume an instantaneous transition from the inflationary to the (RD) stage, the scale factor, the Hubble rate, the mode functions and their (conformal) time derivatives are all \emph{continuous} across the transition and this continuity implies a continuous process of particle production. As a consequence of these continuity conditions the transition \emph{does not} induce a burst of particle production, nor is there any discontinuity in the production dynamics. This important aspect will be highlighted again in sections (\ref{sec:nonad}) and (\ref{sec:tmunu}) below in more detail.

\subsection{Minimal coupling}\label{sec:MC}

We begin by studying the case of minimal coupling (MC), namely $\xi=0$. The mode functions during the inflationary (de Sitter) era are given by (\ref{BDsols2}) for (MC)  and during (RD) the general solution of the mode equations is given by (\ref{generalsol}) in terms of the solutions (\ref{fconf}) with  \emph{out} (particle) boundary conditions.

For the minimally coupled case (MC) we find from eqn. (\ref{BDsols2})
\bea g^<_k(\eta_R) & = & \frac{e^{ik\eta_R}}{\sqrt{2k}}\,\Big[1+\frac{i}{k\,\eta_R} \Big] \label{funmc1}\\ \frac{d}{d\eta}g^<_k(\eta)\Big|_{\eta_R} & = & -ik\,\frac{e^{ik\eta_R}}{\sqrt{2k}}\,\Big[1+\frac{i}{k\,\eta_R}-\frac{1}{(k\,\eta_R)^2} \Big]\label{derfunmc1}\,. \eea Since $k\eta_R \ll 1$ we keep the leading order terms in the superhorizon limit $k\,\eta_R \rightarrow 0$ writing

\bea g^<_k(\eta_R) & = & \frac{i}{\sqrt{2k}\,\delta} \label{funmc}\\ \frac{d}{d\eta}g^<_k(\eta)\Big|_{\eta_R} & = & \,\frac{i\sqrt{k}}{\sqrt{2}\,\delta^2} ~~;~~ \delta = k\,\eta_R \label{derfunmc}\,. \eea From eqn. (\ref{atetaR}) we find
\bea f_k(\eta_R) & = &  \frac{1}{(8mH_R)^{1/4}}\,\Big[\frac{1}{\sqrt{\kappa}}  -i\sqrt{\kappa}  \Big]\,\,W[\alpha;0] \label{fetaR} \\
\frac{d\,f_k(\eta)}{d\eta}\Big|_{\eta_R}  & = &  \frac{\sqrt{2\,m\,H_R}}{(8mH_R)^{1/4}}\,\Big[\frac{1}{\sqrt{\kappa}}  +i\sqrt{\kappa}  \Big]\,\,W'[\alpha;0] \,,\label{defetaR}\eea with
\be W'[a,0] = -\frac{1}{2}\,W[a,0]\,.  \label{idW0}\ee

Using these results with the matching conditions (\ref{ABcoefs}) yield the Bogoliubov coefficients,
\bea A_k & = &  \frac{i}{4\,\delta} \,\Bigg\{ \sqrt{\kappa} \,\Bigg( R_k -\frac{2}{R_k\,\delta} \Bigg) + \frac{i}{\sqrt{\kappa}}   \,\Bigg(  {R_k} +\frac{2}{R_k\,\delta} \Bigg)    \Bigg\}\,,\label{bogAMC}\\
B_k & = &   \frac{i}{4\,\delta} \,\Bigg\{ \sqrt{\kappa} \,\Bigg( R_k -\frac{2}{R_k\,\delta} \Bigg) - \frac{i}{\sqrt{\kappa}}   \,\Bigg(  {R_k} +\frac{2}{R_k\,\delta} \Bigg)    \Bigg\}\,\label{bogBMC} \eea where
\be R_k =     \frac{2^{3/4}}{|\alpha|^{1/4}}  \,
  \Bigg|\frac{\Gamma\Big(\frac{3}{4}- i\frac{|\alpha|}{2}\Big) }{ \Gamma\Big(\frac{1}{4}- i \frac{|\alpha|}{2}\Big)} \Bigg|^{1/2}\,.
\label{Rofk} \ee

Therefore, the distribution function of produced particles is given by
\be \mathcal{N}_k = |B_k|^2 = \frac{1}{4R^2_k \,\delta^4}\,\Bigg[\kappa\,\Big(\frac{R^2_k \delta}{2} -1\Big)^2 +\frac{1}{\kappa}\,\Big(\frac{R^2_k\,\delta}{2} +1\Big)^2\Bigg] \,.\label{distfunmc}\ee It is convenient to extract the relevant scales, hence  define
\be \sqrt{|\alpha|} = \frac{k}{\sqrt{2mH_R}} \equiv z \,,\label{zdef} \ee in terms of which it follows that
\be \delta = k \,\eta_R = z \, \sqrt{\frac{2m}{H_{dS}}} \,,   \label{delz}\ee   yielding
\bea R^2_k\,\delta  & = &   {2^{3/2}}   \,
  \Bigg|\frac{\Gamma\Big(\frac{3}{4}- i\frac{z^2}{2}\Big) }{ \Gamma\Big(\frac{1}{4}- i \frac{z^2}{2}\Big)} \Bigg|\,\, \sqrt{\frac{2m}{H_{dS}}}\nonumber \\  \frac{1}{R^2_k\,\delta^4} & = & \frac{1}{z^3}~~\bigg( \frac{H_{dS}}{m} \bigg)^2 \frac{1}{8\sqrt{2}}\, \Bigg|\frac{\Gamma\Big(\frac{1}{4}- i\frac{z^2}{2}\Big) }{ \Gamma\Big(\frac{3}{4}- i \frac{z^2}{2}\Big)} \Bigg|  \,, \label{facs}\eea with
\be    {\frac{H_{dS}}{m}}= \frac{1}{ {m_{ev}}}~\Bigg[\frac{H_{dS} }{10^{13}\,(\mathrm{GeV})} \Bigg]~~10^{22} \,.\label{delratio}\ee

\begin{figure}[ht!]
\begin{center}
\includegraphics[height=4in,width=4.5in,keepaspectratio=true]{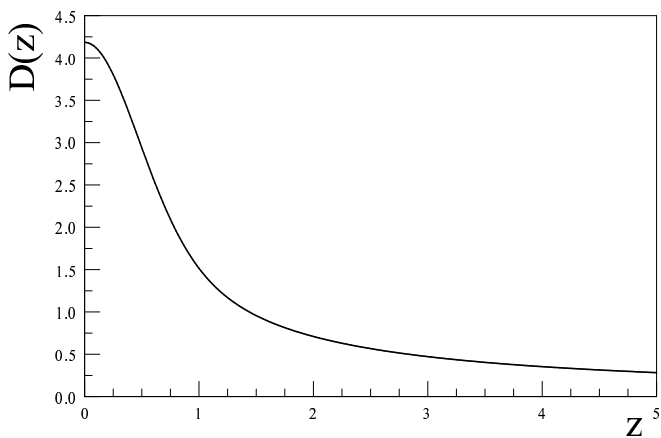}
\caption{ The function D(z) vs. z. }
\label{fig:dofz}
\end{center}
\end{figure}

\begin{figure}[ht!]
\begin{center}
\includegraphics[height=4in,width=4.5in,keepaspectratio=true]{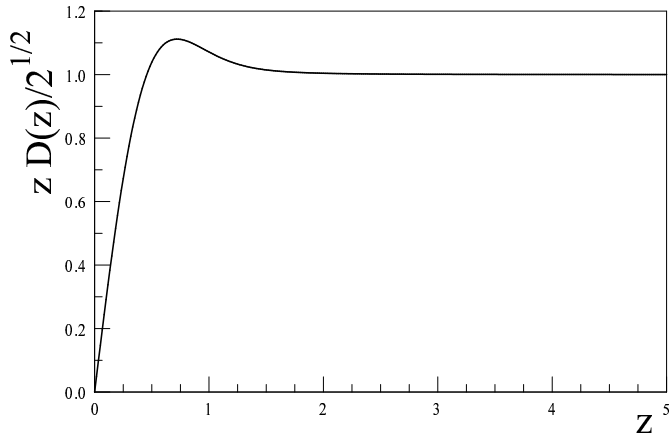}
\caption{ The function $\frac{z}{\sqrt{2}}\,D(z)$ vs. z displaying the asymptotic behavior (\ref{asygams}). }
\label{fig:cdofz}
\end{center}
\end{figure}

Using Stirling's approximation we find that the asymptotic behavior of the ratio of Gamma functions in eqn.(\ref{facs}) is given by
\be  \Bigg|\frac{\Gamma\Big(\frac{1}{4}- i\frac{z^2}{2}\Big) }{ \Gamma\Big(\frac{3}{4}- i \frac{z^2}{2}\Big)} \Bigg| ~~ \overrightarrow{z\rightarrow \infty} ~~ \frac{\sqrt{2}}{z} \,. \label{asygams}\ee We focus  on wavelengths that are superhorizon at the end of inflation, namely $k\eta_R \ll 1$ which results in the following condition
\be k\eta_R = z\,\sqrt{\frac{2m}{H_{dS}}} \ll 1\,. \label{condish}\ee  For large $z$ the product
\be R^2_k\,\delta \rightarrow 2 \,  z\,\sqrt{\frac{2m}{H_{dS}}} = 2 \, k\, \eta_R  \,, \label{prodR2del} \ee therefore in the regime of validity of the superhorizon approximation   $k\,\eta_R \ll 1$, the product $R^2_k\,\delta \ll 1$ and can be safely neglected. Hence we can  approximate the distribution function as
\be \mathcal{N}_k \simeq  \frac{1}{16\sqrt{2}}~~\bigg( \frac{H_{dS}}{m} \bigg)^2 \frac{D(z)}{z^3} \,. \label{nofkmcfin}\ee where
\be D(z)= \sqrt{1+e^{-2\pi z^2}}~~ \Bigg|\frac{\Gamma\Big(\frac{1}{4}- i\frac{z^2}{2}\Big) }{ \Gamma\Big(\frac{3}{4}- i \frac{z^2}{2}\Big)} \Bigg|\,.  \label{Dofz}\ee   Figs. (\ref{fig:dofz}, \ref{fig:cdofz}) display  $D(z) $ and  $zD(z)/\sqrt{2}$ vs $z$  respectively.

The number of produced particles $\mathcal{N}_k$ is strongly peaked at low momentum $\mathcal{N}_k \propto 1/k^3$. This  infrared enhancement     and the factor $H^2_{dS}$ are both  remnants of the infrared behavior of light minimally coupled scalars during the de Sitter era. Because $D(z) \rightarrow \sqrt{2}/z$ for $z \gg 1$ it follows that for large comoving wavevectors $\mathcal{N}_k \rightarrow 1/k^4$. The small and large momentum limits of the distribution function are summarized as follows:
\be \mathcal{N}_k \propto \Bigg\{ \begin{array}{c}
                            1/k^3 ~~;~~ k \ll \sqrt{2mH_R} \\
                             1/k^4 ~~;~~ k \gg \sqrt{2mH_R}
                          \end{array}\,.  \label{nofklims}\ee

 \subsection{Conformal coupling}\label{sec:cc}

 Massless particles conformally coupled to gravity are not affected by the cosmological expansion. Therefore, we expect that very light particles with conformal coupling will not be substantially produced. However, in order to fully compare with the minimally coupled case, we study the production in the conformal case and focus on establishing the main aspects of the difference.

For conformal coupling  the mode functions during the inflationary stage are given by (\ref{BDsols2}) for $\xi = 1/6$.  With  $k\eta_R \ll 1$ we find
\bea g^<_k(\eta_R) & = & \frac{1}{\sqrt{2k}} \label{funcc}\\ \frac{d}{d\eta}g^<_k(\eta)\Big|_{\eta_R} & = & \,\frac{-i\sqrt{k}}{\sqrt{2}}   \label{derfuncc}\,. \eea During the (RD) era the mode functions are given by (\ref{generalsol}) with $f_k(\eta)$ given by (\ref{fconf}). The Bogoliubov coefficients are found in the same manner as for the minimal coupling by equating the functions and $\eta-$ derivatives at $\eta=\eta_R$.

We find
\bea A_k & = &  \frac{1}{4} \,\Bigg\{  \,\Bigg(\sqrt{\kappa}\,R_k + \frac{2}{\sqrt{\kappa}\,R_k} \Bigg) + i   \,\Bigg( \frac{R_k}{\sqrt{\kappa}}+\frac{2\,\sqrt{\kappa} }{R_k} \Bigg)    \Bigg\}\,,\label{bogAcc}\\
B_k & = &  \frac{1}{4} \,\Bigg\{  \,\Bigg(\sqrt{\kappa}\,R_k - \frac{2}{\sqrt{\kappa}\,R_k} \Bigg) - i   \,\Bigg( \frac{R_k}{\sqrt{\kappa}}-\frac{2\,\sqrt{\kappa} }{R_k} \Bigg)    \Bigg\}\,,\label{bogBcc} \eea where $\kappa$ and $R_k$ is given by (\ref{fconf},\ref{Rofk}) respectively. It is straightforward to confirm the identity (\ref{condiAB}). A comparison with the Bogoliubov coefficients of the minimally coupled case, (\ref{bogAMC},\ref{bogBMC}) reveals that $A_k,B_k$ for minimal coupling feature the denominators with $\delta = k\,\eta_R \ll 1$. These denominators are a direct consequence of the infrared enhancement of the mode functions for nearly massless minimally coupled scalar fields in de Sitter space time, as evident in eqns. (\ref{BDsols2}) and (\ref{funmc1},\ref{derfunmc1}).

   The distribution function of produced particles is
\be \mathcal{N}_k = |B_k|^2 = \frac{1}{8} \,\Bigg\{\sqrt{1+e^{-2\pi|\alpha|}}\,\Bigg(R^2_k + \frac{4}{R^2_k} \Bigg) -4  \Bigg\}\,.  \label{nofkcc}\ee

 Using the asymptotic properties of the Gamma functions, we find that  $\mathcal{N}_k \rightarrow 1/(32\alpha^2)^2 \propto 1/k^8$ for $k \rightarrow \infty $ and as $k \rightarrow 0$
\be \mathcal{N}_k \propto \frac{1}{\sqrt{|\alpha|}} \propto \frac{1}{k}\,   \label{ksmall} \ee therefore particles are produced primarily with very small momentum $k \ll \sqrt{m\,H_R}$.

  \begin{figure}[ht!]
\begin{center}
\includegraphics[height=4in,width=4.5in,keepaspectratio=true]{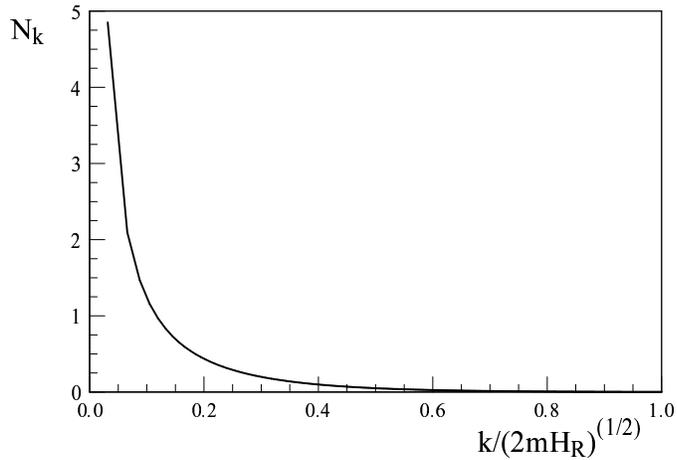}
\caption{ $\mathcal{N}_k$ vs. $z=\sqrt{|\alpha|}= \frac{k}{\sqrt{2mH_R}}$ for conformal coupling. }
\label{fig:nkconformal}
\end{center}
\end{figure}

The distribution function $\mathcal{N}_k$  is solely a function of $z=k/\sqrt{2mH_R}$, fig. (\ref{fig:nkconformal}) displays $N_k$ vs. $z= |\alpha|^{1/2}= k/\sqrt{2mH_R}$. It is then convenient to define the distribution function
 \be \mathcal{N}(z) \equiv \mathcal{N}_k \,,\label{nofzcc}\ee
$\mathcal{N}(z)$  is peaked at low momentum and vanishes fast for $z > 1$, for example $\mathcal{N}(z=1) \simeq 10^{-3}; \mathcal{N}(z=10) \simeq 10^{-7}$. As a corollary, the particles are produced non-relativistically at the time of matter-radiation equality, since
\be \frac{k}{m\,a_{eq}} \lesssim \sqrt{\frac{2H_R}{m\,a^2_{eq}}} \simeq \frac{10^{-13}}{\sqrt{m_{ev}}} \,, \label{nrprod}\ee hence, even for $m_{ev} \simeq 10^{-22}$ it follows that $k/m\,a_{eq} \lesssim 10^{-2}$. Therefore for $m \gtrsim 10^{-22}\,\mathrm{eV}$ the produced particles are non-relativistic at all times after matter-radiation equality.

Although the distribution function is peaked at low momentum, there is a striking difference between the minimal and conformal coupling cases. In the (MC) case $\mathcal{N}_k \simeq 1/k^3$ whereas for (CC) $\mathcal{N}_k \simeq 1/k$ as $k \rightarrow 0$. This difference can be traced to the difference in mode functions during the inflationary stage as displayed by eqn. (\ref{BDsols2}), because during the (RD) era $a''=0$ and the mode equation and mode functions are the same for (MC) and (CC). During the inflationary stage $a''\neq 0$ and minimally coupled fields with masses $m \ll H_{dS}$ feature an infrared enhancement, which propagates through the matching conditions into the Bogoliubov coefficients.

Note that unlike the (MC) case, in the (CC) case the Bogoliubov coefficients $A_k,B_k$ do \emph{not} depend on the scale of inflation $H_{dS}$, this is also a consequence of the infrared enhancement of (MC) light fields during inflation, encoded in the factors $1/k\eta$ in the (MC) mode functions.

During the (RD) era both minimally and conformally coupled fields obey the same equations of motion because $a''=0$ in (RD), hence the mode functions $f_k(\eta)$ are obviously the same in both cases. The difference in behavior for $\eta > \eta_R$ emerges from the different matching conditions with the mode functions during inflation. This leads us to conclude that most of the difference in particle production between these cases is a consequence of the evolution during the inflationary stage.

\section{Non-adiabatic particle production.}\label{sec:nonad}
In the expansion of the field in terms of the exact mode functions (\ref{quantchi}) the annihilation and creation operators $a_{\vk},a^\dagger_{\vk}$ are time independent. Following \cite{parker,birrell,ford,mottola,dunne} we can introduce \emph{time} dependent operators by expanding in the basis of adiabatic  ``out''  particle states. Introduce the zeroth-order adiabatic modes
\be \tf_k(\eta)= \frac{e^{-i\,\int^{\eta}\,\omega_k(\eta')\,d\eta'}}{\sqrt{2\,\omega_k(\eta)}}~~;~~ \omega_k(\eta) = \sqrt{k^2+ m^2\,a^2(\eta)}\,,  \label{zerof}\ee and expand the \emph{exact} mode functions $g_k(\eta)$ as
\be g_k(\eta) = \ta_k(\eta)\,\tf_k(\eta)+ \tb_k(\eta)\,\tf^*_k(\eta) \label{gexpa}\ee and the $\eta-$ derivative (canonical momentum)\cite{parker,mottola,dunne}
\be g'_k(\eta) = Q_k(\eta)\, \ta_k(\eta)\,\tf_k(\eta)+ Q^*_k(\eta) \,\tb_k(\eta)\,\tf^*_k(\eta) \,. \label{dergexpa}\ee  With
\be Q_k(\eta) = -i \omega_k(\eta)+ V_k(\eta) \,,\label{Q}\ee with $V_k(\eta)$ a \emph{real function} it follows that the Wronskian condition (\ref{wronskian}) yields
\be |\ta_k(\eta)|^2-|\tb_k(\eta)|^2 =1\,. \label{Wroab}\ee Inserting the ansatz (\ref{gexpa},\ref{dergexpa}) into the mode equations yields the coupled equations of motion for the coefficients $\ta_k(\eta),\tb_k(\eta)$, obtained in references\cite{mottola,dunne}.
The relations (\ref{gexpa},\ref{dergexpa}) can be inverted to yield the coefficients\cite{mottola}
\bea \ta_k(\eta) & = &  i\,\tf^*_k(\eta) \Big[g'_k(\eta) - Q^*_k(\eta)\,g_k(\eta)  \Big] \label{tilA} \\
\tb_k(\eta) & = &  -i\,\tf_k(\eta) \Big[g'_k(\eta) - Q_k(\eta)\,g_k(\eta)  \Big]\,. \label{tilB}\eea Different choices of the real functions $V_k(\eta)$     yield different dynamics for  coefficients $\ta_k(\eta),\tb_k(\eta)$\cite{mottola,dunne}. Taking, for example $V_k(\eta) =0$ corresponds to the lowest (zeroth) adiabatic order, another choice, $V_k(\eta) = \omega'_k(\eta)/2\omega_k(\eta)$ yields a first adiabatic order correction\cite{mottola,dunne}. For both of these values, the continuity of $a(\eta), H(\eta),g_k(\eta),g'_k(\eta)$ across the inflation to (RD) transition implies the continuity of the coefficients $\ta_k(\eta),\tb_k(\eta)$. Namely   particle production is a continuous process across the transition, and \emph{not} a consequence of the assumption of instantaneous reheating.

The difference in the $\eta$- dependence of the coefficients $\ta,\tb$ for these two choices has been studied in ref.\cite{dunne}. Introducing the expansion (\ref{gexpa}) into (\ref{quantchi}) yields
\be a_{\vk}\,g_k(\eta) + a^\dagger_{-\vk} \, g^*_k (\eta) =c_{\vk}(\eta) \,\tf_k(\eta)+  c^{\dagger}_{-\vk}(\eta)\,\tf^*_k(\eta)\,, \label{adiaexp}\ee where
\be c_{\vk}(\eta) = a_{\vk}\,\ta_k(\eta)+a^\dagger_{-\vk}\,\tb^*_k(\eta)~~;~~c^\dagger_{\vk}(\eta) = a^\dagger_{\vk}\,\ta^*_k(\eta)+a_{-\vk}\,\tb_k(\eta)\,. \label{cops} \ee Therefore the number of \emph{adiabatic} particles at a given time $\eta$ is
\be \widetilde{\mathcal{N}}_k (\eta) = \bra{0}c^\dagger_{\vk}(\eta)\, c_{\vk}(\eta)\ket{0} = |\tb_k(\eta)|^2 \,. \label{adianum}\ee

We now \emph{choose} to expand in the basis of the zeroth-order adiabatic ``out'' particle states, by setting $V_k(\eta)=0$. Note that if $g_k(\eta)$ coincides exactly with the adiabatic mode function $\tf_k(\eta)$ then $\ta_k(\eta)=1;\tb_k(\eta)=0$ and there is no particle production.

During the inflationary stage with $a(\eta) = 1/H_{dS}(\eta-2\eta_R)$ and $\eta < \eta_R$, the mode functions $g_k(\eta)$ are $g^<_k(\eta)$ given by (\ref{BDsols2}). As $\eta \rightarrow -\infty $ ($\eta \ll \eta_R$ ) these approach the adiabatic mode functions $\tf_k(\eta)$, hence it is straightforward to find that
\be \ta_k(\eta) \rightarrow 1 ~~;~~ \tb_k(\eta) \rightarrow 0 \,, \label{initime}\ee yielding as $\eta \rightarrow -\infty$
 \be \widetilde{\mathcal{N}}_k (\eta \rightarrow -\infty) = 0 \,, \label{adiavacuum}\ee namely the initial vacuum state. For super-horizon wavelengths, $k\eta  \ll 1$,  the exact mode functions for minimal coupling (MC) in eqn. (\ref{BDsols2}) differ drastically from the adiabatic ones leading to non-adiabatic particle production when the wavelengths cross the horizon during the inflationary stage.

 During the (RD) stage,   for $\eta > \eta_R$, the mode functions are   $g^{>}_k(\eta)$ given by (\ref{generalsol}) where   $f_k(\eta)$ are solutions of Weber's equations   with ``out'' boundary conditions (\ref{outstates}). At early times after the transition $\eta  \gtrsim \eta_R$, the Weber functions $f_k(\eta)$ differ drastically from $\tf_k(\eta)$, however,  asymptotically at long time $f_k(\eta)$ coincide with   $\tf_k(\eta)$ because of the ``out'' boundary conditions (\ref{outstates}). Therefore, for  $\eta \gg \eta_R$ at asymptotically long time during (RD), it is also straightforward to show that
\be \ta_k(\eta) \rightarrow A_k + \mathcal{O}\big(\omega'_k/\omega^2_k\big) ~~;~~ \tb_k(\eta) \rightarrow B_k + \mathcal{O}\big(\omega'_k/\omega^2_k\big) \,, \label{asyadiard}\ee hence the interpolating time dependent number of particles yields asymptotically during (RD)
\be \widetilde{\mathcal{N}}_k (\eta \gg \eta_R) = |B_k|^2 = \mathcal{N}_k \,. \label{outnumb}\ee This analysis highlights that the ``out'' particles are produced during the time regimes where the exact mode functions depart from the adiabatic ones.   During inflation particle production is    substantially enhanced  after horizon crossing in the minimally coupled case, and continues non-adiabatically into the (RD) era  during the regime of non-adiabatic evolution (\ref{alaful}).    As clearly discussed in ref.\cite{dunne}, different choices of the real function $V_k(\eta)$ yield   \emph{different} time dependence of the interpolating particle number during the non-adiabatic stages, precisely when particles are produced. Nevertheless, the \emph{asymptotic} number of particles coincide with $\mathcal{N}_k$ for \emph{any} definitions of $V_k$ that involve a higher adiabatic ratio\cite{dunne}. For example, choosing $V_k(\eta) = \omega'_k(\eta)/2\omega^2_k(\eta)$ as in ref.\cite{mottola}, the asymptotic in and out behavior as $\eta \rightarrow -\infty$  and $\eta \gg \eta_R$ remain the same because the adiabatic ratio vanishes in the asymptotic limits. Therefore, whereas the definition of particles and the evolution of the time dependent interpolating particle number depends on the particular choice of basis vectors (adiabatic order) and  the  real function $V_k(\eta)$,  the \emph{aymptotic} (out) particle number $\mathcal{N}_k$ is \emph{independent} of such choice.

During inflation, for a minimally coupled  light scalar field the mode functions are \emph{not} adiabatic after the corresponding wavelength becomes superhorizon, namely as $k \eta\ll 1$ as evidenced by the exact mode functions for the (MC) case given by eqn. (\ref{BDsols2}). As we have stated above, during the (RD) era after inflation, the Weber mode functions are also non-adiabatic after the transition for superhorizon wavelengths. The production of ``out'' particles occurs primarily during the non-adiabatic evolution and is \emph{continuous} across the transition from inflation to (RD) domination. As discussed above, this is a consequence of the continuity of scale factor, Hubble rate, mode functions and their conformal time derivative across the transition.

For a conformally coupled (CC) light particle, and with $m/H_{dS} \ll 1$ the mode function in the inflationary era, given by eqn.  (\ref{BDsols2}) (CC), does not differ substantially from $\tf_k(\eta)$, hence there is very little production during the inflationary era, unlike the minimally coupled case. Hence we expect, that the (CC) case will yield a much smaller abundance, an expectation that is confirmed by the analysis of the energy momentum tensor below. Furthermore, during (RD) both minimally and conformally coupled fields obey the same equations of motion, while the corresponding mode functions are drastically different during inflation. Therefore,  the difference in the evolution for $\eta > \eta_R$ between these cases is  imprinted from the inflationary stage through the matching conditions.

 While there is a quantitative difference in the dynamics for different choices of $V_k(\eta)$, the above  statements remain  true for any choice consistent with the adiabatic expansion, as demonstrated in the study of ref.\cite{dunne}.  Furthermore, regardless of the precise definition of an interpolating time dependent particle number, ultimately what is needed to understand the production of dark matter and its cosmological impact is the energy momentum tensor associated with the (ULDM) field.

\section{The energy momentum tensor: renormalization, abundance and equation of state}\label{sec:tmunu} The energy momentum tensor for the real scalar field $\phi(x)$ with generic coupling to gravity is given by
\bea T_{\mu \nu}  &  =   & (1-2\xi)\,\phi_{,\mu}\phi_{,\nu}-\frac{1}{2}\,(1-4\xi) g^{\alpha \beta}\,\phi_{,\alpha}\phi_{,\beta}\,g_{\mu \nu}-2\xi\,\phi\,\phi_{;\mu ; \nu}+\frac{1}{2}\,(1-6\xi)m^2\,\phi^2\,g_{\mu \nu} \nonumber \\
& + & \frac{\xi}{2}\,g_{\mu \nu} \phi\,\square \phi-\xi\,\Big[R_{\mu \nu} -\frac{1}{2}\,(1-6\xi)\,R g_{\mu \nu} \Big]\phi^2 \,. \label{Tmunu} \eea

 Writing $\phi(x)$ in terms of the conformally rescaled field $\chi(x)$ as in eqn. (\ref{rescaledfields}) and with the mode expansion (\ref{quantchi}) the expectation value of the energy momentum tensor in the Bunch-Davies vacuum state defined by eqn. (\ref{BDvac}) in the spatially flat FRW cosmology is given by\footnote{We take the infinite volume limit with $\frac{1}{V}\sum_{\vk} \rightarrow \int \frac{d^3k}{(2\pi)^3}$. }
 \bea  \bra{0}T{^0}_0\ket{0} = \rho(\eta)  &  = &  \frac{1}{4\pi^2\,a^4(\eta)}\,\int^{\infty}_0  k^2 dk \,\Bigg\{ |g^{'}_k(\eta)|^2+  \omega^2_k(\eta)\, |g_k(\eta)|^2 \nonumber \\ &  - &  \,(1-6\xi)\Bigg[ \frac{a'}{a}\,\Bigg(g_k(\eta)g^{'*}_k(\eta)+g^{'}_k(\eta)g^{*}_k(\eta) -\frac{a'}{a} |g_k(\eta)|^2 \Bigg) \Bigg\} \,,\label{tmunuofg}
 \eea
 \bea \bra{0}T{^\mu}_\mu\ket{0} &  =  & \rho(\eta)-3 P(\eta)   =    \frac{1}{2\pi^2\,a^4(\eta)}\,\int^{\infty}_0  k^2 dk \,\Bigg\{m^2\,a^2(\eta)\,|g_k(\eta)|^2 \nonumber \\ & - &  (1-6\xi)\,\Bigg[|g^{'}_k(\eta)|^2 -
 \omega^2_k(\eta)\,|g_k(\eta)|^2-
 \frac{a'(\eta)}{a(\eta)}\,\Bigg(g_k(\eta)g^{'*}_k(\eta)+g^{'}_k(\eta)g^{*}_k(\eta)  \Bigg) \nonumber \\ & - & \Bigg( \frac{a''(\eta)}{a(\eta)}-\Big(\frac{a'(\eta)}{a(\eta)}\Big)^2\Bigg) |g_k(\eta)|^2   +(1-6\xi)\,|g_k(\eta)|^2\,\frac{a''(\eta)}{a(\eta)}\Bigg] \Bigg\}\,, \label{traceofg}
 \eea
 where $\rho(\eta),P(\eta)$ are the energy density and pressure respectively. Using the mode equations (\ref{chimodes}) it is straightforward to show the covariant conservation of  $\bra{0}T{^\mu}_\nu\ket{0}$.  We note that the continuity of the scale factor, the Hubble rate and the mode functions and their conformal time derivatives at the inflation-(RD) transition  at $\eta_R$ guarantees the continuity of the energy density $\bra{0}T{^0}_0\ket{0}$ as is evident from eqn. (\ref{tmunuofg}). Hence particle production is \emph{not} a consequence of the approximation of a sudden transition but rather a consequence of the non-adiabatic evolution, as emphasized previously.

 The instantaneous reheating approximation, with the continuity of mode functions, scale factor and Hubble rate across the transition, cannot yield a continuity in $a''$. The reason for this is physically clear: the expectation value of the energy momentum tensor \emph{of the background}  in the homogeneous and isotropic Bunch-Davies vacuum is of the ideal fluid form $\bra{0}T{^\mu}_\nu\ket{0} = \mathrm{diag}(\rho, -P, -P, -P)$ with $\bra{0}T{^\mu}_\mu\ket{0} = \rho - 3 P$. The Ricci scalar $R = 6 a''/a^3 \propto \bra{0}T{^\mu}_\mu\ket{0} = \rho - 3 P$, during the inflationary stage the equation of state is $P = -\rho$ yielding $\bra{0}T{^\mu}_\mu\ket{0} \neq 0$ whereas in an (RD) era $P = \rho/3$ and  $\bra{0}T{^\mu}_\mu\ket{0} = 0$ hence a vanishing Ricci scalar\footnote{This neglects the conformal anomaly\cite{bunch,anderson}.}. Therefore instantaneous reheating implies a discontinuity in the Ricci scalar, hence $a''$.  For the scalar (DM) particle  $ \bra{0}T{^\mu}_\mu\ket{0}$ given by (\ref{traceofg})  depends explicitly on $a''$, therefore, while the energy density is continuous, the pressure features a discontinuity as a consequence of the change in the background equation of state for instantaneous reheating.

 During the inflationary stage $\eta < \eta_R$ the mode functions are $g^{<}_{k}(\tau)$ given by (\ref{BDsolution} ) corresponding to the ``in'' Bunch-Davies vacuum state. Therefore during this stage the energy density is simply the \emph{zero point energy density} associated with the Bunch-Davies vacuum.

 For $\eta > \eta_R$, the mode functions in (\ref{tmunuofg},\ref{traceofg}) are $g_k(\eta) =g^>_k(\eta)= A_k\,f_k(\eta) + B_k\,f^*_k(\eta)$,   with the Bogoliubov coefficients given by eqns. (\ref{ABcoefs}) obeying the relation (\ref{condiAB}). We now write  $\bra{0}T^{\mu}_{\nu}\ket{0}$ in terms of the mode functions $f_k(\eta)$ describing the asymptotic particle states with ``out'' boundary conditions. Since we are interested in the energy momentum tensor near matter radiation equality we average   over rapidly varying phases  in the interference terms of the form $f f, f^* f^*$ (and derivatives). We find
 \bea  \bra{0}T{^0}_0\ket{0} = \rho(\eta)  &  = &  \frac{1}{4\pi^2\,a^4(\eta)}\,\int^{\infty}_0  k^2 dk \,\Big(1 + 2\,\mathcal{N}_k\Big)\,\Bigg\{ |f^{'}_k(\eta)|^2+  \omega^2_k(\eta)\, |f_k(\eta)|^2 \nonumber \\ &  - &  \,(1-6\xi)\Bigg[ \frac{a'}{a}\,\Bigg(f_k(\eta)f^{'*}_k(\eta)+f^{'}_k(\eta)f^{*}_k(\eta) -\frac{a'}{a} |f_k(\eta)|^2 \Bigg) \Bigg\}\,, \label{tmunuoff}
 \eea
 \bea \bra{0}T{^\mu}_\mu\ket{0} &  =  &   \frac{1}{2\pi^2\,a^4(\eta)}\,\int^{\infty}_0  k^2 dk \,\Big(1 + 2\,\mathcal{N}_k\Big)\,\Bigg\{m^2\,a^2(\eta)\,|f_k(\eta)|^2 \nonumber \\ & - &  (1-6\xi)\,\Bigg[|f^{'}_k(\eta)|^2 -
 \omega^2_k(\eta)\,|f_k(\eta)|^2-
 \frac{a'(\eta)}{a(\eta)}\,\Bigg(f_k(\eta)f^{'*}_k(\eta)+f^{'}_k(\eta)f^{*}_k(\eta)  \Bigg) \nonumber \\ & - & \Bigg( \frac{a''(\eta)}{a(\eta)}-\Big(\frac{a'(\eta)}{a(\eta)}\Big)^2\Bigg) |f_k(\eta)|^2   +(1-6\xi)\,|f_k(\eta)|^2\,\frac{a''(\eta)}{a(\eta)}\Bigg] \Bigg\}\,,
 \eea
  where $\mathcal{N}_k = |B_k|^2 $ and used the relation (\ref{condiAB}). The next step consists of expanding $W_k(\eta)$ defining the WKB form of the mode functions (\ref{WKB}) in the  adiabatic expansion (\ref{adexp}). We follow the steps in ref.\cite{bunch,anderson,mottola} and expand the expectation values of the energy momentum tensor up to fourth order in the adiabatic expansion, with the result
  \bea \rho(\eta)  & = &  \rho^{(0)}(\eta) + \rho^{(2)}(\eta)+ \rho^{(4)}(\eta)+\cdots   \label{rhoad} \\
  \bra{0}T{^\mu}_\mu\ket{0} & = &  \mathcal{T}^{(0)}(\eta)+\mathcal{T}^{(2)}(\eta)+\mathcal{T}^{(4)}(\eta)+\cdots \label{tracead}   \eea where the superscripts refer to the order in the adiabatic expansion. The respective contributions are similar to  the results of ref.(\cite{bunch,anderson}) but with the extra factor $1+2\mathcal{N}_k$ in the integrand.

  Of  particular interest for this study are the zeroth adiabatic order energy density and pressure, which are given by
  \bea \rho^{(0)}(\eta) & = &  \frac{1}{4\pi^2\,a^4(\eta)} \int^{\infty}_0 k^2 \,[1+2\mathcal{N}_k]\,\omega_k(\eta)\, dk \,, \label{rhozero} \\
  \mathcal{T}^{(0)}(\eta) & = & \frac{1}{4\pi^2\,a^4(\eta)} \int^{\infty}_0 k^2 \,[1+2\mathcal{N}_k]\,\frac{m^2 \,a^2(\eta)}{\omega_k(\eta)}\, dk \,, \label{Tzero}
  \eea yielding
  \be P^{(0)}(\eta) = \frac{1}{3}\,\Big[\rho^{(0)}(\eta) - \mathcal{T}^{(0)}(\eta)\Big] = \frac{1}{12\pi^2\,a^{4}(\eta)} \,\int^{\infty}_0 [1+2\mathcal{N}_k]\,\frac{k^4}{\omega_k(\eta)}\, dk \,.\label{Pzero} \ee

  The energy momentum tensor features ultraviolet divergences that must be regularized and renormalized. This is explicit at zeroth adiabatic order given by eqns. (\ref{rhozero},\ref{Tzero}), the higher order adiabatic corrections can be found in ref.\cite{bunch,anderson} by multiplying the integrand in momentum by the factor $1+2\mathcal{N}_k$. Appendix (\ref{app:emt}) shows some second order adiabatic contributions that   yield ultraviolet divergences in $\langle T^{\mu}_{\nu}\rangle$ for $\mathcal{N}_k=0$. These adiabatic terms feature inverse powers of $\omega_k$ as befits the adiabatic expansion, in particular $1/\omega_k;1/\omega^3_k$ which yield quadratic and logarithmic ultraviolet divergences.

  For the minimally coupled case $\mathcal{N}_k \propto k^{-4}$ at large momenta (see eqn. (\ref{nofklims})) . Therefore the terms with $\mathcal{N}_k$ for the higher adiabatic orders  do  not contribute to the ultraviolet divergences.  Consider for example the second adiabatic corrections $\rho^{(2)}$, explicitly given in   appendix (\ref{app:emt}), as compared to the zeroth order contribution during the radiation dominated area near matter radiation equality (\ref{rhozero}) it is suppressed  by a factor
  \be \propto \Big(\frac{a'}{ma}\Big)^2 \simeq   \Big(\frac{H_R}{m\,a_{eq}}\Big)^2 \simeq \Bigg(\frac{10^{-31}}{m_{ev}}\Bigg)^2\,, \label{supfac}\ee with much larger suppression factors for the terms of higher adiabatic order. The same argument holds for $\mathcal{T}^{(2)}$, for which several contributions are explicitly given in appendix (\ref{app:emt}).

  \vspace{1mm}

  \textbf{Renormalization:}

  The ultraviolet divergences in the energy momentum tensor must be regularized and renormalized. For $\mathcal{N}_k =0$ such a program is well established and has been thoroughly studied and implemented in refs.\cite{birrell,pf,fh,hu,bunch,anderson,mottola,bir}. As discussed in detail in these references, the ultraviolet divergences are absorbed into renormalizations of the cosmological constant, Newton's constant $G$ and into the geometric tensors $H^{(1,2)}_{\mu \nu}$ which result from the variational derivative of a gravitational action that includes higher curvature terms $\propto R^2, R^{\mu\nu} R_{\mu \nu}$. These higher curvature terms are added in the action multiplied by counterterms, which are then required to cancel the coefficients of the geometric tensors in such a way that the renormalized action is the Einstein-Hilbert action.

  Since our focus is to study the contribution from particle production, namely $\mathcal{N}_k\neq0$ we absorb the \emph{full} energy momentum tensor for $\mathcal{N}_k = 0$ into these renormalizations, this is tantamount to subtracting the \emph{zero point or vacuum energy density} during the inflation and radiation eras. After this subtraction and renormalization, only the terms proportional to $\mathcal{N}_k$ in (\ref{rhozero}) and (\ref{tracead}) are considered.

  Since, as shown explicitly in eqn. (\ref{nofklims})   $\mathcal{N}_k \propto 1/k^4$ as $k\rightarrow \infty$ for the minimally coupled case, the   corrections of second adiabatic order   and higher \emph{do not feature} ultraviolet divergences, and are suppressed by factors of order $10^{-62}/m^2_{ev}$ near matter-radiation equality. Hence,  we keep solely the contribution of zeroth adiabatic order from particle production. After renormalization and  to leading adiabatic order we find the contributions to the energy density and pressure from particle production to be given by

  \bea \rho^{(pp)}(\eta) & = &  \frac{1}{2\pi^2\,a^4(\eta)} \int^{\infty}_0 k^2 \, \mathcal{N}_k \,\omega_k(\eta)\, dk \,, \label{rhozeropp} \\
  P^{(pp)}(\eta) & = &  \frac{1}{2\pi^2\,a^{4}(\eta)} \,\int^{\infty}_0   \,\frac{1}{3}\, k\,v_k(\eta)\,\mathcal{N}_k \, k^2  dk  ~~;~~ v_k(\eta) = \frac{k}{\omega_k(\eta)}\,.\label{Pzeropp}  \eea This result is noteworthy: the density and pressure are \emph{exactly} the diagonal components of a \emph{kinetic} energy momentum tensor describing a (perfect) fluid. Note that the integrals are over comoving momentum, in terms of the physical (local) energy $E_k(\eta) = \sqrt{k^2_{ph}(\eta) + m^2}$ and physical momenta $k_{ph}(\eta) = k/a(\eta)$ these expressions can be written as

 \bea \rho^{(pp)}(\eta) & = &  \frac{1}{2\pi^2} \int^{\infty}_0   \mathcal{F}\big[a(\eta)\,k_{ph}\big] \,E_k(\eta)\,k^2_{ph}\, dk_{ph} \,, \label{rhozeropp2} \\
  P^{(pp)}(\eta) & = &  \frac{1}{2\pi^2 } \,\int^{\infty}_0   \,\frac{1}{3}\, k_{ph}\,\frac{k_{ph}}{E_k(\eta) }\,\mathcal{F}\big[a(\eta)\,k_{ph}\big]  \, k^2_{ph}  dk_{ph}  \,,\label{Pzeropp2}  \eea where
  \be \mathcal{F}\big[a(\eta)\,k_{ph}\big] \equiv \mathcal{N}_k \,,\label{dist} \ee is a \emph{frozen}, i.e. a time independent  distribution function of produced particles.  It is straightforward to show covariant conservation, namely
  \be \dot{\rho}^{(pp)}(t) + 3 \,\frac{\dot{a}(t)}{a(t)} \,\Big(\rho^{(pp)}(t) + P^{(pp)}(t)\Big) =0    \,. \label{covaconspp}\ee

  We highlight this result: \emph{the usual fluid-kinetic energy momentum tensor emerges as the leading order (zeroth order) in the adiabatic expansion after subtracting the ``vacuum'' contribution which is absorbed in the renormalization of the cosmological and Newton's constant, and cancel counterterms that multiply higher curvature terms in the action}.  The full expectation value of the energy momentum tensor during the non-adiabatic stage \emph{cannot} be written   in the kinetic form in terms of the distribution function; such simplification is \emph{only} available during adiabatic evolution.

  As discussed above, in the minimally coupled case  the distribution function $\mathcal{N}_k \propto 1/k^4$ in the large $k$ limit, therefore   both, the energy density (\ref{rhozeropp}) and pressure (\ref{Pzeropp}) at zeroth adiabatic order  feature \emph{a priori} ultraviolet logarithmic divergences.  However, these divergences are actually beyond the realm of validity of \emph{two} of our main approximations,  \textbf{i:)} superhorizon wavelengths at the end of inflation, namely $k\eta_R \ll1$.     As discussed in section (\ref{sec:asy}), taking the upper bound on the scale of inflation  this condition   implies that $k \ll \mu\mathrm{eV}$, this is hardly an ultraviolet large cutoff in momentum.  Therefore,  in principle and  for consistency, the momentum integrals must be cutoff at this scale, thus the ``divergences'' associated with particle production are not physical. \textbf{ii:)}  as discussed in detail in section (\ref{sec:discussion}) the assumption of instantaneous reheating will definitely not be warranted for sub-horizon wavelengths,  and the distribution function for these (large) wavevectors (with $k \gg \mu\mathrm{eV}$) \emph{may} differ drastically from that of the wavevectors that are super-Hubble at the end of inflation. Hence, consistency with our  main assumptions imply that the contributions from particle production to the energy momentum tensor must be cut-off at a comoving momentum scale $\simeq \sqrt{H_R\,H_{dS}} \simeq \mu\mathrm{eV}$ for $H_{dS} \simeq 10^{13}\,\mathrm{GeV}$, which corresponds to wavelengths longer than   a meter.

  Therefore, we regularize the integrals featuring $\mathcal{N}_k$ by  introducing a comoving upper momentum cutoff $k_{max} \lesssim 1/\eta_R = \sqrt{H_R\,H_{dS}}$.   Because the distribution function $\mathcal{N}_k$ is enhanced at low momentum    we also include a \emph{lower momentum cutoff} $k_{min} \simeq H_0$ corresponding to horizon-sized wavelengths     \emph{today}.   Hence the energy density and pressure from particle production are given by

 \bea \rho^{(pp)}(\eta) & = &  \frac{1}{2\pi^2\,a^4(\eta)} \int^{k_{max}}_{k_{min}}  \mathcal{N}_k \,\omega_k(\eta)\,k^2 \,  dk \,, \,, \label{rhozeroppfin} \\
  P^{(pp)}(\eta) & = &  \frac{1}{6\pi^2\,a^{4}(\eta)} \,\int^{k_{max}}_{k_{min}} \mathcal{N}_k    \frac{k^2}{\omega_k(\eta)}\, \, k^2  dk \,.\label{Pzeroppfin}  \eea The abundance $\Omega(a)$ and the equation of state $w(a)$ are, respectively,
  \be \Omega(a) = \frac{\rho^{(pp)}(\eta)}{\rho_c} ~~;~~ \rho_c = \frac{3\,H^2_0}{8\pi\,G}\simeq 0.4\,\times 10^{-10}\, (\mathrm{eV})^4 \,,\label{abundance}\ee
\be w(a) = \frac{P^{(pp)}(\eta)}{\rho^{(pp)}(\eta)}\,. \label{eqnofstate} \ee

\vspace{1mm}

\subsection{Minimal coupling:}

For the minimal coupling case $\mathcal{N}_k$ is given by eqn. (\ref{nofkmcfin}) in terms of the variable $z$ defined by eqn. (\ref{zdef}), in this case we find the abundance (\ref{abundance})

\be \Omega(a) = \frac{m}{\rho_c\, a^3(\eta)}\,\,\bigg( \frac{H_{dS}}{m} \bigg)^2\,\Big( m\,H_R\Big)^{3/2}\, \frac{1}{16\,\pi^2}\,    \int^{z_M}_{z_m} D(z)~\Bigg[\frac{z^2}{a^2(\eta)}\, \Big(\frac{2 H_R}{m}\Big) +1  \Bigg]^{1/2}\,\frac{dz}{z}\,.\label{Omabmc} \ee The minimum $z_m$ provides an infrared cutoff, with $k_{min} \simeq H_0$, it follows that  $z_m = H_0/\sqrt{2mH_R}$. Values of comoving momentum $k \gg m$ inside the integral of the distribution function yield contributions that redshift as $1/a^4(\eta)$ hence contributing to the radiation component.  The matter contribution for $a(\eta)  \gtrsim a_{eq}$ is extracted from contributions to the integrals from  comoving momenta  $k \lesssim m\,a_{eq}$, hence we introduce an upper cutoff $z_M \leq m a_{eq}/\sqrt{2mH_R}$.  Therefore for $a(\eta) > a_{eq}$ we find the contribution to the (DM) abundance
\be \Omega(a) \simeq 0.5\, \,\frac{\sqrt{m_{ev}}}{a^3(\eta)}\,\,\Big[\frac{H_{dS}}{10^{13}\,\mathrm{GeV}} \Big]^2\,\, \int^{z_M}_{z_m}\frac{ D(z)}{z}~ \, {dz} \equiv \frac{\Omega_{pp}}{a^3(\eta)}\,.\label{abumc2}\ee Taking as the maximum comoving wavevector $k \simeq m \,a_{eq}$  and the minimum $k \simeq H_0$ it follows that $z_{M} \simeq \sqrt{m_{ev}}\,\times 10^{13} \gg 1$  and $z_{m} \simeq H_0/\sqrt{2\,m\,H_R} \simeq 10^{-16}/\sqrt{m_{ev}} \ll 1$, hence    $D(z_{M}) \simeq 10^{-13}/\sqrt{m_{ev}} \ll 1$ and $D(z_{m}) \simeq \sqrt{2}\, \frac{\Gamma(\frac{1}{4})}{\Gamma(\frac{3}{4})}$.  Upon integration by parts the integral in (\ref{abumc2}) is given by
\be \int^{z_M}_{z_m}\frac{ D(z)}{z}~ \, {dz} \simeq    -\sqrt{2}\, \frac{\Gamma(\frac{1}{4})}{\Gamma(\frac{3}{4})}\,\ln\Big[ \frac{H_0}{\sqrt{2\,m \, H_R}}\Big]  -  \underbrace{ \int^{\infty}_{0}\ln(z) \frac{d\,D(z)}{dz}  ~ {dz}}_{\simeq 0.6}\,,  \label{intabun}\ee where in the second term (integral) we have taken $z_{m} \rightarrow 0; z_{M} \rightarrow \infty$ because the integrand vanishes fast at both limits, and the remaining integral is carried out numerically. Therefore to  leading order we find
\be \Omega(a) = \frac{\Omega_{pp}}{a^3(\eta)}~~;~~ \Omega_{pp} = 2.09\, \sqrt{m_{ev}}\,\Big[\frac{H_{dS}}{10^{13}\,\mathrm{GeV}} \Big]^2\,\,\ln\Big[ \frac{\sqrt{2\,m \, H_R}}{H_0}\Big] \,.\label{ODM}\ee

For a given value of $m_{ev}$ this equation yields the contribution to the dark matter abundance as a function of $m_{ev}$ and the \emph{only} uncertain cosmological parameter $H_{dS}$.  Requiring that the  abundance $\Omega_{pp} =\Omega_{DM}=0.25$   gives the dependence of the mass that yields the correct abundance on  $H_{dS}$, namely
 \be \sqrt{m_{ev}}\,\Big[\ln\big[\sqrt{m_{ev}} \big]+ 36 \Big] = 0.12 \Bigg[\frac{10^{13}\mathrm{GeV}}{H_{dS}} \Bigg]^2\,. \label{mofHds}\ee For $H_{dS} \simeq 10^{13}\,\mathrm{GeV}$ we find that the correct (DM) abundance yields the value

\be m  \simeq 1.5 \times 10^{-5}  \, \mathrm{eV}
\,. \label{massboundmc}\ee

\begin{figure}[ht!]
\begin{center}
\includegraphics[height=5in,width=4.5in,keepaspectratio=true]{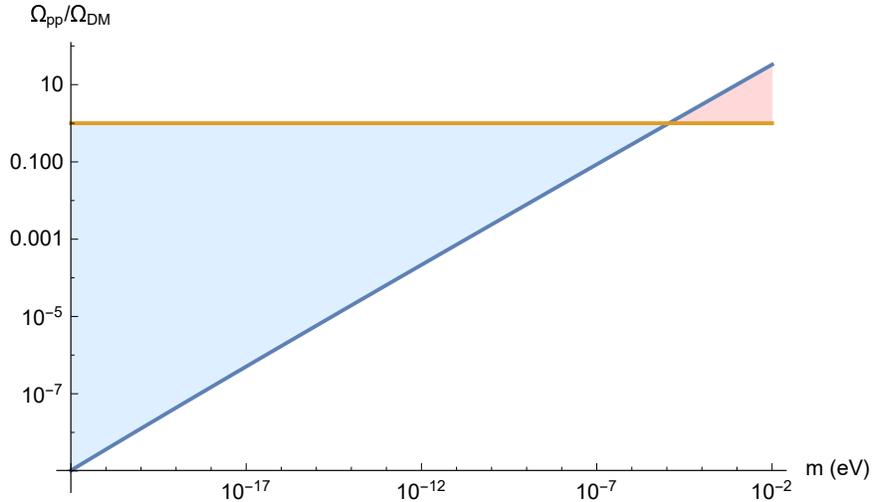}
\caption{ $\ln\Big[\frac{\Omega_{pp}}{\Omega_{DM}}\Big]$ vs. $\ln[m_{ev}]$ for $H_{dS}=10^{13}\,\mathrm{GeV}$. The blue-shaded region corresponds to under abundance and the red-shaded to overabundance (colors online).  }
\label{fig:omega}
\end{center}
\end{figure}

The   super-horizon approximation $k\,\eta_R\ll1$ entails a maximum value of the mass for which the approximations involved are consistent. We have set the maximum value of the momentum integral as $k_{M}\simeq m\,a_{eq}$ so as to capture all the values of momenta that contribute to the (non-relativistic) matter contribution. For this upper limit to be consistent with the superhorizon approximation it follows that the mass of the (ULDM) particle is constrained by the upper limit
\be m\,a_{eq}\,\eta_R \lesssim 1  \Rightarrow m \lesssim 0.02\, \, \Bigg[\frac{H_{dS}}{10^{13}\mathrm{GeV}} \Bigg]^{1/2}\,\,\mathrm{eV}\,.\label{ubm} \ee Fig. (\ref{fig:omega}) displays $\ln\Big[\frac{\Omega_{pp}}{\Omega{DM}}\Big]$ with $\Omega_{DM}=0.25$ vs. $\ln[m_{ev}]$ for $H_{dS}=10^{13}\,\mathrm{GeV}$.

The pressure and equation of state are given by eqns. (\ref{Pzeroppfin},\ref{eqnofstate}) respectively. For the non-relativistic component describing a matter dominated ``fluid'' we take $\omega_k(\eta) = m\,a(\eta)$ in the integrands. The remaining integrals are   similarly obtained   with the above cutoffs. The equation of state parameter is given by
\be w(a) = \frac{2}{3}\frac{H_R}{m\,a^2(\eta)} ~ \frac{\int^{z_M}_{z_m}  D(z)\,z\,dz}{\int^{z_M}_{z_m}\frac{ D(z)}{z}\,dz}\,,\label{eqnofstatemc} \ee taking  $z_{M} = m a_{eq}/\sqrt{2mH_R}$ and $z_m \simeq H_0/\sqrt{2mH_R}$ we find
\be w(a) \simeq \frac{2\,\Gamma(3/4)}{3\,\Gamma(1/4)}\, \frac{\Big(\frac{H_R}{2ma^2_{eq}}\Big)^{1/2}}{\ln\Big[\frac{\sqrt{2mH_R}}{H_0} \Big] }\, \Big( \frac{a_{eq}}{a(\eta)}\Big)^2\,. \label{wofadm}\ee Taking the value of the mass as given by (\ref{massboundmc}) with $H_{dS} \simeq 10^{13}\,\mathrm{GeV}$ we find
\be w(a_{eq}) \simeq 2.5 \times 10^{-14}\,.  \label{waeq}\ee For a non-relativistic species we find \be   \langle V^2(\eta) \rangle = \frac{\int \mathcal{N}_k\, \frac{k^2}{m^2 a^2(\eta)}\,k^2 dk }{\int \mathcal{N}_k\,  k^2 dk } \equiv 3\frac{P(\eta)}{\rho(\eta)} = 3 \,w(a)\,.  \label{v2} \ee

Therefore, indeed this is a very cold dark matter candidate despite being so light. The main reason is that the distribution function strongly peaks at small values of momentum. The redshift behavior of $w(a)$ is that expected for a non-relativistic component.

 \vspace{1mm}

 \textbf{Free streaming length:}

 The comoving free streaming wave-vector is defined in analogy with the Jeans wavevector in the fluid description of perturbations, namely\cite{wu}
 \be k^2_{fs}(\eta) = \frac{4\pi\, G\,\rho_m(\eta)}{\langle V^2(\eta) \rangle}\,a^2(\eta) = \frac{3}{2} \, \frac{ H^2_0\,\Omega_m}{\langle V^2_{eq} \rangle\,a^2_{eq}}\, a(\eta) \,, \label{k2fs}\ee where $\langle V^2(\eta) \rangle$ is given by eqn. (\ref{v2}), which we have written as
 \be \langle V^2(\eta) \rangle = \langle V^2_{eq} \rangle \,\Big(\frac{a_{eq}}{a(\eta)}\Big)^2\,. \label{v2eq} \ee   As shown in ref.\cite{wu} the cutoff scale in the power spectrum is the comoving free streaming length
 \be \lambda_{fs} \equiv \frac{2\pi}{k_{fs}(a_{eq})} = 2\pi \, \Big[\frac{2\,\langle V^2_{eq} \rangle\,a_{eq} }{3\,\Omega_M} \Big]^{1/2}~  d_H  \,, \label{lfs} \ee where $d_H = 1/H_0 = 3\,\mathrm{Gpc}/h$ is the Hubble distance. This definition differs from the usual definition of the comoving free streaming distance $l_{fs}$ during matter domination by factors of $\mathcal{O}(1)$:
 \be l_{fs} = \int^{\eta_0}_{\eta_{eq}} \sqrt{\langle V^2(\eta) \rangle}\,d\eta = \sqrt{\langle V^2_{eq} \rangle}~a_{eq} \,\int^{\eta_0}_{\eta_{eq}} \frac{ d\eta}{a(\eta)} \,.\label{llfs}\ee During the matter dominated era it follows that
 \be d\eta = \frac{1}{H_0\,\sqrt{\Omega_M}}\,\frac{da}{a^{1/2}}\,, \label{detarel}\ee hence the free streaming distance from matter-radiation equality until $a_0 \simeq \mathcal{O}(1)$ is given by
 \be l_{fs} = 2 \, \Big[\frac{ \langle V^2_{eq} \rangle\,a_{eq} }{ \Omega_M} \Big]^{1/2}~  d_H \,. \label{lfsfinal}\ee   Using the results (\ref{waeq},\ref{v2})) corresponding to $H_{dS} \simeq 10^{13}\,\mathrm{GeV}$, we find
 \be \lambda_{fs} \simeq l_{fs} \simeq 70\,\mathrm{pc}\,. \label{fsfinal}\ee This is the cutoff scale in the matter power spectrum; thus we see that even for a very light (DM) candidate with $m \simeq 10^{-5} \,\mathrm{eV}$ the cosmological production yields a very \emph{cold} species with a rather
 small free streaming length comparable to that of heavy weakly interacting massive particles.

\subsection{Conformal coupling}

For the case of conformal coupling, the distribution function $\mathcal{N}_k$  that enters in the abundance and equation of state (\ref{rhozeroppfin}-\ref{eqnofstate}) is given by (\ref{nofkcc}).
 The integral for the density, eqn. (\ref{rhozeroppfin}), cannot be obtained in closed form. However,   $\mathcal{N}_k $ is solely a function of $z =k/\sqrt{2mH_R}$ and localized in the region $0 \leq z \leq 1$ as discussed in section (\ref{sec:cc}) and displayed in fig. (\ref{fig:nkconformal}). Furthermore for $a(\eta) \simeq a_{eq}$ this region of comoving momenta correspond to non-relativistic particles and we can safely replace $\omega_k(\eta) \simeq m\,a(\eta)$ inside the integrand in  (\ref{rhozeroppfin}),  yielding
(near matter radiation equality)
\be \rho^{(pp)}(\eta)   =   \frac{m}{ a^3(\eta)} \int   \mathcal{N}_k  \,k^2 \,  \frac{dk }{2\pi^2} \,,   \label{rhozeroppNR} \ee therefore the low momentum peak of the distribution function entails that the density redshifts as non-relativistic matter.

Changing variables to $z$  and writing $\mathcal{N}_k \equiv \mathcal{N}(z)$, we find
\be \rho^{(pp)}(\eta)   =  \frac{1}{2\pi^2} \frac{m^4}{ a^3(\eta)}\,\Big[\frac{2\,H_R}{m} \Big]^{3/2} \int^{z_M}_0   \mathcal{N}(z)  \,z^2 \,   {dz }  \,,   \label{rhozeroppNRcc} \ee where $z_M \lesssim m \,a_{eq}/\sqrt{2mH_R}$ and the lower limit can be taken to zero because the integrand does not feature an infrared divergence.

The remaining integral is rapidly convergent and is carried out numerically with an upper limit $z\simeq 20$ (with $\mathcal{N}(20) \simeq 10^{-20}$), for which the integral yields the value $\simeq 0.01$. Hence  we find   the abundance
\be \Omega(a) \simeq 1.3\times \frac{1}{a^3(\eta)}\,\Big[\frac{m}{ (\mathrm{eV})} \Big]^4\, \,\Big[\frac{2H_R}{m}\Big]^{3/2} \, \times 10^{7} \simeq \frac{(m_{ev})^{5/2}}{a^3(\eta)}\times 10^{-46}  \,. \label{rhoeq} \ee Thus,  even for $m\simeq (\mathrm{eV})$ the  dark matter abundance for  conformally coupled particles    is negligible. This is in qualitative agreement with our expectations of very small abundance in this case, but implementing the framework described in the previous section allowed us to obtain a quantitative understanding of the abundance in this case.

The main differences with the minimally coupled case can be traced back to the factors $\delta = k\eta_R$ in eqns. (\ref{bogAMC},\ref{bogAMC}). These are a result of the behavior $\propto 1/(k\eta_R)$ of the (MC) mode functions during the inflationary stage (see eqn. (\ref{BDsols2}), a hallmark of the infrared enhancement of correlations of nearly massless particles minimally coupled to gravity in de Sitter space time. These factors result in the infrared enhancement $\mathcal{N}_k \propto 1/k^3$ and the factor $H^2_{dS}$ for the (MC) case vs. $\mathcal{N}_k \propto 1/k$ for the (CC) case.

\section{On entropy perturbations:}\label{sec:iso}

 Adiabatic and entropy perturbations from inflation have been thoroughly studied in refs.\cite{gordon,byrnes,bartolo}, and we refer the readers to these for details.
 In  ref.\cite{gordon,byrnes} the case of two fields is studied in detail; this is the case that is most relevant for our discussion: one of the fields is the inflaton,  the other is the (ULDM) field with action given by (\ref{lagrads}). While the inflaton field develops an expectation value that drives the inflationary stage, the (ULDM) \emph{does not} acquire an expectation value, and is taken to be in its Bunch-Davies vacuum state. In ref.\cite{gordon} ``adiabatic'' and ``entropy'' fields are obtained from the \emph{fluctuations} of the two fields around their expectation value, by introducing a ``mixing'' angle that depends explicitly on the time derivative of the \emph{expectation values} of both fields. The ``adiabatic'' field represents a fluctuation along the background trajectory, while the ``entropy'' field is the orthogonal combination in terms of the ``mixing'' angle. We identify the (ULDM) field with the second ($\chi$)-field in ref.\cite{gordon}. Since in our case this field \emph{does not} acquire an expectation value, it follows that the ``mixing'' angle vanishes identically. In this case, the inflaton fluctuations are the ``adiabatic'' field  and the   (ULDM) field is identified with the ``entropy'' field. Therefore, considering perturbations  \emph{linear in the fluctuations},   the vanishing of the mixing angle implies that the entropy perturbation \emph{does not} source the long-wavelength evolution of the comoving curvature perturbation, nor is there any cross correlation between the adiabatic and entropy perturbations (see for example eqns.(47,48,52,55) and comment below eqn. (50) in ref.\cite{gordon}) .

  Ref.\cite{sena}  focused on superheavy dark matter, and  following on previous study in ref.\cite{mazu}  recognized that in the case in which the dark matter field does \emph{not} acquire an expectation value the treatment of isocurvature perturbations must be modified substantially. The authors of ref.\cite{sena} also recognized that when the superheavy dark matter field does not acquire an expectation value (background) there is no mixing between the fluctuations of this and the inflaton field to linear order\footnote{See the discussion prior to eqn. (87) in ref.\cite{sena}.}.

  The treatment advocated in  ref.\cite{sena} \emph{defines} the energy density perturbation of the dark matter field as
 \bea \delta \rho^{(dm)}(\vec{x}) & = &  \frac{:T^{(dm)}_{00}(\vec{x}): - \langle :T^{(dm)}_{00}(\vec{x}):\rangle}{\rho^{(dm)}} \label{delsena} \\ {\rho^{(dm)}} & = & \langle :T^{(dm)}_{00}(\vec{x}):\rangle\,,  \label{rhosena}\eea where normal ordering is referred to the Bogoliubov rotated vacuum state (see ref.\cite{sena} for details), and identifies the power spectrum of entropy perturbations from the spatial  Fourier transform of  the connected correlation function, namely
 \be \int \frac{d^3r}{(2\pi)^3} ~ e^{i\vec{k}\cdot\vec{r}}\, \langle \delta \rho^{(dm)}(\vec{x}) ~\delta \rho^{(dm)}(\vec{x}+\vec{r}) \rangle \propto \mathcal{P}^{(dm)}(k)\,.  \label{Powdm}\ee In free field theory the connected correlator in eqn. (\ref{Powdm}) is a one loop diagram.

  In   ref.\cite{sena} the expectation value $\rho^{(dm)}$ depends explicitly on the Bogoliubov coefficient $\beta$ associated with particle production during inflation, and \emph{vanishes identically} when this coefficient vanishes, which is the case in our study.

 The above  definitions \emph{do not} apply to our case since  during the inflationary stage the quantum state in our treatment is the Bunch-Davies vacuum state and consequently,   the energy momentum tensor during this stage describes the \emph{zero point} energy density of this vacuum state. There is no Bogoliubov coefficient $\beta$ and as per the result eqn. (90) in ref.\cite{sena} the energy density $\rho^{(dm)}$ given by (\ref{rhosena})  \emph{vanishes identically}. Furthermore, as discussed in section (\ref{sec:tmunu}) we have renormalized the energy momentum tensor by subtracting the full contribution from the \emph{zero point energy density} during inflation and the radiation eras.  Therefore the definition (\ref{delsena}) cannot be applicable to our study.

 There is another important caveat in the interpretation of entropy perturbations advocated in ref.\cite{sena}: as we discussed in detail in section (\ref{sec:tmunu}) the expectation value of the energy momentum tensor features quartic, quadratic and logarithmic divergences and requires subtractions up to fourth adiabatic order to be renormalized. These aspects had already been addressed in references\cite{bunch,parker,anderson,hu}. The various divergences are absorbed into renormalizations of the cosmological and Newton constants, but also in higher curvature counterterms in the bare action (corresponding to the tensors $H^{(1.2)}_{\mu \nu}$ \cite{bunch}).

 Different regularizations (subtractions) yield different finite contributions to the energy density, therefore  the finite contribution to the expectation value yielding $\rho^{(dm)}$ is not uniquely defined and depends on the subtraction scheme. As discussed in section (\ref{sec:tmunu}) we substract the \emph{full zero point energy} all throughout the evolution. In fact this procedure subtracts completely the zero point contribution during inflation and radiation eras, hence in our case $\rho^{(dm)}$ \emph{vanishes identically} during inflation after renormalization. This is the usual procedure in semiclassical gravity, for example during inflation only the background (expectation value) contribution is considered and in radiation domination only the finite temperature (kinetic) contributions to the energy momentum tensor are considered. Furthermore, there are several other caveats associated with the definition of the power spectrum (\ref{Powdm}) proposed in ref.\cite{sena} (see eqn. (96) in ref.\cite{sena}): \textbf{i:)} it is straightforward to show that the kinetic term contribution to the energy momentum tensor yields an ultraviolet divergence with the \emph{fifth power} of an ultraviolet cutoff to $\mathcal{P}^{(dm)}(k)$. In ref.\cite{sena} this divergence in the kinetic contribution is not manifest because this contribution is evaluated near the end of inflation when the mode functions are dominated by the superhorizon contributions and the integrals have been cutoff in the ultraviolet, with an upper momentum $a_e H_e$.  But the leading ultraviolet divergences are similar to those in Minkowski space time and dominate the earlier dynamics.    \textbf{ii:)} even  for the mass term contribution of the energy momentum tensor, the one-loop connected diagram that yields $\mathcal{P}^{(dm)}(k)$ still features a linear  ultraviolet divergence, which was neglected in ref.\cite{sena} because it is multiplied by a function of  time that becomes vanishingly small near the end of inflation. \textbf{iii:)} unlike the renormalization of the energy momentum tensor whose subtractions are absorbed systematically into the parameters of the total action (including higher curvature terms), there is no natural manner to absorb the divergences in the power spectrum (\ref{Powdm}). In ref.\cite{sena} all  the divergent integrals are cutoff at wavevectors $\simeq a_e \,H_e$, namely those that cross the horizon at the end of inflation. However, a complete treatment should include a proper renormalization of the divergences and the zero point energy. In our study, the fact that during inflation the \emph{full} energy density is the zero point corresponding to the Bunch-Davies vacuum makes the framework to describe non-linear entropy perturbations advocated in ref.\cite{sena}  not applicable to our case. Some of these caveats have been recognized in ref.\cite{sena}\footnote{See for example the comments after eqn.  (86)  in ref.\cite{sena}.}.

 \vspace{1mm}

 %%% new addition  #1 %%%%

 \textbf{Entropy perturbations post-inflation:} The discussion above has focused on the generation of entropy perturbations \emph{during inflation} and the applicability of the framework introduced in ref.\cite{sena}. However, the important aspect is the impact of entropy (isocurvature) perturbations upon the (CMB). In the usual approach to cosmological perturbations, adiabatic and isocurvature perturbations during inflation provide the initial conditions of the respective perturbations upon horizon re-entry during the radiation (or matter) dominated era. As  discussed in detail in  refs.\cite{bartolo,byrnes}, the initial conditions of isorcurvature perturbations are determined by  the set of transfer functions discussed in ref.\cite{byrnes}. These, in turn, are proportional to the ``mixing'' (or correlation) angle associated with the expectation value of the entropy field   (see for example eqns. (44, 45) in \cite{bartolo}), which in our case \emph{vanishes identically}.  Furthermore, the framework introduced in ref.\cite{sena} does not apply to our case as discussed above. Therefore, for the case that we study, the initial conditions for isocurvature perturbations during the radiation dominated era \emph{cannot} be determined in the inflationary stage. As discussed above,  the energy momentum tensor during inflation describes the vacuum zero point energy and is completely subtracted out by renormalization. In the post-inflationary stage it features three contributions: the vacuum contribution is subtracted out in the renormalization procedure, the interference term is rapidly oscillating in the adiabatic regime and therefore its expectation value averages out on short time scales, and   the contribution from particle production, which in the adiabatic regime features the kinetic fluid form. It is this latter term that is the relevant one (after renormalization) to understand dark matter perturbations, the distribution function is completely determined by the Bogoliubov coefficient $|B_k|^2$. The influence of isocurvature perturbations on the (CMB) is a result of solving the system of Einstein-Boltzmann equations for \emph{linear} cosmological   perturbations, in which  $|B_k|^2$ is the distribution function of the \emph{unperturbed} (DM) component, and   $\rho_{pp}$ (\ref{rhozeroppfin}) describes the \emph{background density}. This set of Einstein- Boltzmann equations  must be appended with initial conditions, which are   determined from the respective super-horizon perturbations at the end of inflation. From the above discussion, it is clear that in the case that we study,  the proper initial conditions for isocurvature perturbations remain to be understood at a deeper level.

 %%% end of new addition #1 %%%%

 \vspace{1mm}

 The corollary of this discussion is that a proper definition of the power spectrum of entropy perturbations in the case when the fields do \emph{not} acquire expectation values remains to be understood at a   deeper level. The caveats associated with the renormalization of the energy momentum tensor along with its correlations remain to be clarified in a consistent and unambiguous manner. These include a proper account of the fact that there is no natural manner to renormalize the divergences in a power spectrum obtained from the connected correlation function of the energy momentum tensor. These remain even when the zero point contribution to the energy density is completely subtracted. The contribution of zero point energy correlations to non-linear perturbations merits deeper scrutiny, \textit{since  even the fluctuations of the inflaton yield zero point contributions to the energy density and all other fields that are either produced or excited post-inflation   presumably  {also} contribute to the zero point energy density during inflation}.  A satisfactory resolution of these important issues, necessary   to   quantify reliably the impact of \emph{non-linear} entropy perturbations is still lacking,  and is clearly well beyond the scope of this study.

\section{Discussion and caveats.}\label{sec:discussion}

\textbf{On reheating:} Reheating dynamics, namely the non-equilibrium processes that lead to a (RD) dominated era after the inflationary stage are still being vigorously studied. Most studies of reheating necessarily input particular forms for the inflaton potential and model the couplings of (standard model) particles to the inflaton and/or other degrees of freedom thereby yielding model dependent descriptions with widely different time scales depending on unknown couplings and masses\cite{reheat}.

One of our main assumptions is that the transition from the inflationary stage to the (RD) dominated stage is instantaneous. The main physical reason behind this approximation is that we focus on wavelengths that are superhorizon at the end of inflation. The dynamics of the mode functions for these wave-vectors is on long time scales, hence insensitive to the reheating dynamics occurring on much shorter time scales. Furthermore, in principle, wavelengths larger than the particle horizon are causally disconnected from the causal microphysical processes of thermalization. While this assumption \emph{seems} physically reasonable, it must be tested quantitatively. However, this requires studying a particular model of reheating dynamics. While conclusions of a particular model will not be universally valid, perhaps a simple model that dynamically and \emph{continuously} interpolates (with continuous scale factor and Hubble rate) between a near de Sitter inflationary stage and a post-inflation (RD) stage would illuminate the validity of the instantaneous approximation. Most likely such study would require a substantial numerical effort to solve the mode equations during the transition and matching to the solutions in the subsequent (RD) era. Clearly such study is beyond the scope of this article but merits further attention.

\textbf{Inflationary particle production:} During the (RD) era the equations of motion are the same for (MC) and (CC)    fields because $a''(\eta)=0$. However, during inflation the equations of motion for the two cases are very different, yielding the drastically different solutions given by eqn. (\ref{BDsols2}). Whereas the mode functions for the (CC) case are ``close'' to the adiabatic mode functions, those of the (MC) depart substantially when the wavelength becomes superhorizon $k\eta \ll 1$. This difference is imprinted on the evolution of the mode functions for $\eta > \eta_R$ through the matching conditions (continuity of function and derivative at $\eta_R$).  The results from the (CC) case confirm negligible particle production in this case, this leads us to conclude that the largest contribution to particle production in the (MC) case occurs during the inflationary stage. This conclusion is bolstered by the analysis of section (\ref{sec:nonad}), where it is shown that the (MC) mode functions depart substantially from the adiabatic ones   for superhorizon modes thus resulting in substantial particle production, whereas those for the (CC) case are similar to the adiabatic ones with little particle production.

\textbf{  Bose Einstein condensate vs. distribution function :}  We have shown that for minimally coupled ultra-light particles the \emph{distribution function} peaks at \emph{very low} comoving momentum with $\mathcal{N}_k \propto 1/k^3$. As discussed in the previous section the distribution function of the produced particles ``inherits'' the infrared enhancement of  the mode functions of minimally coupled ultra-light particles during the inflationary era (taken to be a de Sitter space-time). This enhancement, however, does \emph{not} imply Bose Einstein condensation, particle number of the real scalar field is not conserved, and the field does not acquire a vacuum expectation value. Namely, there is no off-diagonal long range order and no expectation value that would break a $U(1)$ symmetry both of which are  typically associated with Bose-Einstein condensation.    The description of this (ULDM) is in terms of the contributions to the energy momentum tensor. This is very different from the phenomenological Schroedinger-Poisson equation advocated for ``fuzzy'' dark matter\cite{fuzzyDM,fuzzy2,wittenfuzzy} which relies on a ``many-body'' Schroedinger-like wave function for a classical order parameter field akin to the Gross-Pitaevskii equation (non-linear Schroedinger equation) for a superfluid. In many body physics such equation is typically obtained from a variational derivative of the  expectation value of a many-body Hamiltonian in a coherent state\cite{langer}.

\textbf{Self-consistency and backreaction:} We have taken the cosmological expansion as a (RD)  background, neglecting the contribution of the (ULDM) to the radiation component. Such contribution is obtained from the  momentum region with $k \gg m\,a_{eq}$ in the integrals for the density and pressure. In principle this contribution modifies the ulrarelativistic content of the plasma contributing a term that redshifts like radiation $\propto 1/a^4(\eta)$ and, in principle, should be treated self-consistently. However, we consider that the (RD) era is dominated by the $\simeq 100$  ultrarelativistic degrees of freedom of the standard model (and possibly beyond), therefore the contribution of one extra degree of freedom, can be neglected as a first approximation.

\textbf{Lower bound on abundance:} Including possible interactions with either the inflaton or other fields within or beyond the standard model entails \emph{additional} production mechanisms for a very long lived (DM) particle.  Production from reheating or from other mechanisms only \emph{increases} the abundance, and loss mechanisms, such as decay, will occur on time scales comparable to or larger than the Hubble time today. Therefore, this study yields a \emph{baseline} for the production of ultra-light dark matter particles; any other production mechanism will increase the abundance. This is an important corollary of our study: this simplest of models describing the darkest of dark matter (only gravitational interactions) yields an abundance from non-adiabatic particle production which must be accounted for in any model of (ULDM) particles featuring interactions. Thus the abundance resulting from this mechanism is a lower bound to the abundance of \emph{any interacting} species of long-lived (ULDM), and applies, for example to axion-like candidates.

\textbf{Similarities and differences with vector dark matter production:} the production of a massive vector boson during inflation has been recently studied in ref.\cite{graham}. The authors show that the longitudinal component behaves similarly to a massive, minimally coupled scalar field, whereas the transverse components are conformally coupled to gravity. Remarkably, in this reference it is found that the abundance of the longitudinal component is very similar to the result eqn. (\ref{ODM}) above (up to logarithmic contributions). While the equivalence between the longitudinal component and a massive, minimally coupled scalar field is, perhaps expected, the origin of the similarity in the abundance is by no means clear to us.

 In particular we match the mode functions with ``in'' (Bunch-Davies vacuum)  boundary conditions during inflation to the \emph{exact}  mode equations in the radiation dominated era with ``out'' boundary conditions determined by the positive adiabatic frequencies at long time near matter radiation equality, with the matching conditions described in section (\ref{sec:asy}). Furthermore we obtain the full   energy momentum tensor, confirm covariant conservation  and identify the particle production contribution after the proper renormalization and well into the adiabatic regime when the renormalized energy momentum tensor attains the kinetic form in terms of the distribution function. Finally, the total matter density is obtained from the integral of this distribution function, which again is extracted during the adiabatic regime. While perhaps all of these aspects are somehow included in the scaling argument in ref.\cite{graham}, we have not been able to find the proper equivalence between our treatment and the framework of ref.\cite{graham}. However, this aspect notwithstanding, the similarities between the abundance in both results is remarkable.

\textbf{Caveats:} The result (\ref{mofHds}) implies that for  a  very low inflation scale, namely with $H_{dS} \ll 10^{13}\,\mathrm{GeV}$ and  for a fixed, given mass $m_{ev}$ the (ULDM) gravitationally produced yields a much smaller abundance. Or, equivalently,  the value of the mass that yields the correct (DM) abundance increases substantially, whereas consistency of the approach requires the upper bound given by (\ref{ubm}). Since there is a large uncertainty on the scale of inflation, to be resolved by a clear measurement of primordial gravitational waves (or the tensor-to-scalar ratio), it is possible that a very low scale would lead to a revision of the assumption on instantaneous reheating.  Furthermore, the only direct observational evidence of a (RD) era is from Big Bang Nucleosynthesis via the primordial abundance of light elements; this scale, however, corresponds to a few $\mathrm{MeV}$. Thus it is possible that the reheating temperature is as low as a few $\mathrm{MeV}$\cite{lowT}. If this  were the case, a very large discrepancy between the scale of inflation and the reheating temperature cannot be accommodated within the instantaneous reheating approximation because modes that are superhorizon during inflation may re-enter during the dynamical evolution between the end of inflation and the (RD) stage, thus modifying the final distribution function even for long wavelengths. Such large discrepancy   will require a fundamental understanding of the cosmological evolution \emph{between} the two eras suggesting that there may be  a long epoch after the end of inflation that is \emph{not} described by a (RD) cosmology. This scenario  would invalidate one of our main assumptions and require a completely different approach to describing   cosmological production, and at the fundamental level, a complete revision of assumptions on post-inflationary cosmology.

\section{Conclusions.}\label{sec:conclusions}

We have studied the non-adiabatic cosmological production of ultra-light dark matter particles under a minimal set of assumptions: a single ultra-light real scalar field that \emph{only} interacts with gravity and no other field, it is a spectator field in its Bunch-Davies vacuum state during inflation, it does not contribute to the inflationary dynamics nor to any linear metric perturbation (such as isocurvature). We focus on superhorizon wavelengths after inflation, since these are the cosmologically relevant scales for structure formation,  and assume an instantaneous reheating into a (RD) cosmology. The cases of minimal and conformal coupling to gravity are analyzed separately. The mode equations in either case are solved \emph{exactly} both in the inflationary and the (RD) eras with a continuous matching of scale factor, Hubble rate, mode functions and conformal time derivative at the transition. These  continuity conditions imply the continuity of the energy density across the transition. The ``out'' particle states are carefully defined in terms of the zeroth-order adiabatic states at asymptotically long time after the transition, these states are \emph{locally} identified with particle states as in Minkowski space-time. The matching conditions at the transition between inflation and (RD) yield the Bogoliubov coefficients from which we obtain the \emph{distribution function} of produced particles. We establish a correspondence with a (conformal) time dependent particle number by introducing an adiabatic basis of ``out'' particle states and show explicitly that particle production is a direct consequence of \emph{non-adiabatic} cosmological evolution during inflation and well into the (RD) era. We show that for a mass $10^{-22}\,\mathrm{eV} \lesssim m $ cosmological evolution becomes \emph{adiabatic} well before matter-radiation equality.  The number of produced particles \emph{only} depends on cosmological parameters. Whereas a conformally coupled light scalar particle is produced with negligible abundance, there is substantial production for minimally coupled light particles with masses much smaller than the Hubble scale during inflation.  The distribution function of minimally coupled light fields feature an infrared enhancement ``inherited'' from the inflationary stage yielding a behavior  $\mathcal{N}_k \propto 1/k^3$ at small comoving wavevectors.  We obtain the full energy momentum tensor for the (ULDM) from which we obtain the energy density and pressure  near matter-radiation equality after renormalization, which is performed by subtracting the zero point energy density during inflation and radiation domination. An important result is that the fully renormalized energy momentum tensor coincides with the fluid-kinetic one at zeroth-order in the adiabatic expansion. The abundance and equation of state depend solely on the mass and cosmological parameters, in particular the scale of inflation for the minimally coupled case. The  main results of this study are the following, for a minimally coupled (ULDM): the ratio of the abundance of produced particles $\Omega_{pp}$ to $\Omega_{DM}$ is given by

\be \frac{\Omega_{pp}}{\Omega_{DM}} = 8.36 \, \Bigg\{\sqrt{m_{ev}}\,\Big[\ln\big[\sqrt{m_{ev}} \big]+ 36 \Big] \Bigg\}\, \Bigg[\frac{H_{dS}}{10^{13}\mathrm{GeV}} \Bigg]^2 \nonumber \ee where $m_{ev} = m/(eV)$ and $H_{dS}$ the Hubble scale during inflation. For the upper bound on the scale of inflation from Planck\cite{planck2018} $H_{dS} \simeq 10^{13}\,\mathrm{GeV}$,   we find that the produced particles saturate the (DM) abundance for
\be m \simeq 1.5\, \times 10^{-5}\,\mathrm{eV}\,.  \nonumber\ee  For this value of the mass we find the equation of state parameter at matter-radiation equality
\be w(a_{eq}) \simeq 2.5 \times 10^{-14} \,, \nonumber  \ee and a free streaming length (cutoff scale of the matter power spectrum)

\be \lambda_{fs}   \simeq 70\,\mathrm{pc}\,. \ee Therefore the produced particles while very light are a \emph{cold} dark matter candidate with a free streaming length comparable to that of weakly interacting massive particles.

This   is the simplest model for the darkest of (ULDM) since this particle only features gravitational interactions. As such, the results for the abundance provide a \emph{lower bound} and a baseline for the abundance of any (ULDM) candidate  with a lifetime equal to or longer than $1/H_0$.  Interactions with degrees of freedom of the standard model or beyond that leads to particle production will only \emph{increase} the abundance. This lower bound applies to axion-like particles and must be accounted for in the (DM) contribution of \emph{any} (ULDM) candidate.
A study of cosmological production of fermionic degrees of freedom will be reported elsewhere\cite{herfer}.

We have also discussed the caveats associated with a proper treatment of isocurvature perturbations in the case when the (ULDM) (entropy) field does not acquire an expectation value, suggesting that a deeper understanding of this case is needed for a reliable estimate of isocurvature perturbations from the (ULDM) field. We note that such an analysis has not been done even for the inflaton fluctuations.

\appendix

\section{Connection between the mode functions (\ref{fconf}) and WKB asymptotics.}\label{app:weber}

The mode functions (\ref{fconf}) can be written as
\be f_k(\eta) = \frac{|\mathcal{F}(x,\alpha)|}{(8mH_R)^{1/4}}~e^{-i\varphi(x,\alpha)}\,, \label{modphase} \ee  with $x,\alpha$ defined in eqn. (\ref{weberparas}).  For $|\alpha| \gg x^2$ the Weber function features the asymptotic behavior\cite{as}
\bea |\mathcal{F}(x,\alpha)| &  = &  \frac{1}{|\alpha|^{1/4}}\,\Big[1- \frac{x^2}{16|\alpha|} +\cdots \Big]  = \frac{(2mH_R)^{1/4}}{\sqrt{k}} \,\Big[1- \frac{1}{4} \,m^2 \,H^2_R\,\eta^2 +\cdots  \Big]  \nonumber \\ \varphi(x,\alpha) & = & \frac{\pi}{4} + \sqrt{|\alpha|}\,x\,\Big[1+\frac{2\,x^2}{48\,|\alpha|} +\cdots \Big] = \frac{\pi}{4}+ k\,\eta\,\Big[1+ \frac{m^2\,H^2_R\,\eta^2}{6\,k^2} +\cdots \Big]\,. \label{largea} \eea   And for $x^2 \gg |\alpha|$
\bea |\mathcal{F}(x,\alpha)| &  = & \frac{\sqrt{2}}{\sqrt{x}} \,\Big[1- \frac{|\alpha|}{x^2} +\cdots   \Big] = \frac{\sqrt{2}}{(2mH_R)^{1/4}\,\sqrt{\eta}} \,\Big[1- \frac{k^2}{4\,m^2\,H^2_R\,\eta^2} +\cdots \Big]\nonumber  \\ \varphi(x,\alpha) & = & \frac{x^2}{4}+|\alpha|\,\ln(x) + \cdots = \frac{1}{2} m H_R\,\eta^2 + \frac{k^2}{2mH_R}\,\ln[\eta\,\sqrt{2mH_R}] + \cdots \label{largex} \eea
 Up to an overall constant phase these expansions coincide with the expansions of
 \be f_k(\eta) = \frac{e^{-i\int^{\eta} \omega_k(\eta')\,d\eta'}}{\sqrt{2\omega_k(\eta)}} \label{fasy} \ee in both limits $k \gg  m\,H_R\,\eta$ and $k \ll  m\,H_R\,\eta$ respectively.

\section{Second order adiabatic contributions to $T_{\mu \nu}$}\label{app:emt}
We gather the results of second adiabatic order  for the expectation value of the energy momentum tensor (see \cite{anderson,bunch}).

\be \rho^{(2)}(\eta) =   \Big[\frac{a'}{m\,a} \Big]^2  \, \frac{1}{4\pi^2\,a^4(\eta)}\,\int^{\infty}_0  k^2 dk \,m\,\Big(1 + 2\,\mathcal{N}_k\Big)\, \Big[\frac{m^5\,a^4}{8\omega^5_k}+ \frac{1}{2}\, (1-6\xi)\,\Big(\frac{m}{\omega_k}+\frac{m^3 a^2}{\omega^3_k}\Big)\Big]\,  \,. \label{rho2}  \ee

\bea && \mathcal{T}^{(2)}(\eta)   =   \frac{1}{4\pi^2\,a^4(\eta)}\,\int^{\infty}_0  k^2 dk  \,\Big(1 + 2\,\mathcal{N}_k\Big)\, \Bigg\{\frac{m^6\,a^4}{4\,\omega^5_k}\,\Bigg[\frac{a''}{m^2\,a}+\Big(\frac{a'}{m\,a} \Big)^2 \Bigg]-\frac{5\,m^8\,a^6}{8\,\omega^7_k} \,\Big(\frac{a'}{m\,a}\Big)^2 +\nonumber \\ & + &  (1-6\xi) \,\Bigg[\frac{m^2}{\omega_k}\,\Bigg(\frac{a''}{m^2\,a}-\Big(\frac{a^{'}}{m\,a} \Big)^2 \Bigg)+\frac{m^4 a^2}{2\,\omega^3_k}\,\Bigg(2\frac{a''}{m^2\,a}-\Big(\frac{a}{m\,a}\Big)^2 \Bigg)-\frac{3\,m^6\,a^4}{\omega^5_k}\,\Big(\frac{a'}{m\,a}\Big)^2  \Bigg]  \Bigg\} \,.\label{T2} \eea

The   terms with $1/\omega_k;1/\omega^3_k$ yield ultraviolet divergences for $\mathcal{N}_k=0$, which are subtracted and absorbed into the renormalization counterterms as discussed in section (\ref{sec:tmunu}),  whereas the term proportional to $\mathcal{N}_k$ yields ultraviolet finite contributions because $\mathcal{N}_k \lesssim 1/k^4 $ at large $k$. During the (RD) dominated era and near matter-radiation equality, these terms are suppressed by a factor
\be \simeq \Big(\frac{a'}{m\,a}\Big)^2 \simeq \frac{10^{-62}}{m^2_{ev}}\,, \label{supp}\ee with respect to the zeroth-adiabatic order contributions (\ref{Tzero},\ref{Pzero}). A similar analysis confirms that the terms of fourth adiabatic order which feature $\mathcal{N}_k$ in the integrand  are much further suppressed and can be safely neglected.

\end{document}